\documentclass[aps,prl,onecolumn,superscriptaddress,groupedaddress]{revtex4}
\usepackage{subfig}
\usepackage{graphicx}  
\usepackage{dcolumn}   
\usepackage{bm}        
\usepackage{amssymb}   
\usepackage{slashed}
\usepackage{graphicx}				
\usepackage{amsmath}
\usepackage{mathtools}
\usepackage{tikz,pgf}
\usepackage{comment}
\usepackage{asymptote}
\usetikzlibrary{arrows,backgrounds}
\usetikzlibrary{fit,scopes,calc,matrix,positioning,decorations.pathmorphing}
\usepackage[all]{xy}
\usepackage{yfonts}
\newcommand{\bra}[1]{\ensuremath{\left\langle#1\right|}}
\newcommand{\ket}[1]{\ensuremath{\left|#1\right\rangle}}
\newcommand{\Bracket}[1]{\ensuremath{\left\langle#1\right\rangle}}

\begin{document}
\title{Quantum correlation engineering and the causal structure}
\author{Andrei T. Patrascu}
\address{FAST Foundation, Destin FL, 32541, USA\\
email: andrei.patrascu.11@alumni.ucl.ac.uk}
\begin{abstract}
Starting with the assumption that propagation of classical light determines the causal structure of spacetime and the assumption that the causal structure may be emergent from quantum correlations I show a method through which the causal structure can be engineered by the manipulation of carefully chosen ancillas that allow the manipulation of properties of classical light by means of a quantum entanglement structure obtained using an ancilla purification of classical light. The application of Uhlmann gauge and the creation of a dynamical Uhlmann gauge theory enables us to access the quantum informational structure that determines the classical propagation speed of light in vacuum, and hence gives us a dictionary that allows us to modify the quantum entanglement structure that determines the causal structure. As such, a modification of the causal structure is obtained from purely quantum informational principles, without the need for any "exotic" matter or other impossible types, geometries, or distributions of matter inside a relativistic stress energy tensor. In fact, it is shown that the quantum information tools necessary to produce measurable modifications of the causal structure are experimentally accessible today in a quantum optics or quantum sensors lab. 
\end{abstract}
\maketitle
\section{Introduction}
Einstein's special relativity has prescribed the existence of a maximal speed at which causality and interactions can propagate. This maximal speed, due to its absolute nature with respect to choices of reference frames became the origin for the construction of spacetime as we know it. The resulting causal structure became an unescapable construct in all scientific endeavours related to flat, Minkowski spacetime. Harmonisation of causality with the desire of some researchers to obtain apparent higher speeds than the speed of light, at least as measured by distant observers have always met practical difficulties due to the requirement for strange forms of matter (for example the so called Alcubierre drive [1]). The impractical nature of such explorations originates in the desire to apply the Einstein field equations in their usual form and therefore to engineer the stress energy tensor within them. This is obviously extremely difficult, as the normal requirements for the creation of Alcubierre bubbles of spacetime would require types of matter that are not to be expected in nature [2] and geometries that are unlikely to be realised practically [3].
I propose here an alternative that would have a similar result, but with a radically different origin. Instead of explicitly altering the stress energy tensor in Einstein's field equation, my approach is to modify the causal structure and implicitly to change the maximal speed of the universe (the speed of light) by direct manipulations of ancillas in a quantum optics experiment. The result would be to obtain a localised region of spacetime in which the causal structure is altered in the sense that the light cones are slightly extended and the effective speed of light is increased. 
As the modification occurs at the level of a quantum coherence space, as defined in what follows, the alteration occurs in a universal way, with a direct impact on the causal structure of spacetime, if we are to assume that the intrinsic quantum correlations are what determines the causal structure of spacetime in the first place. I make here this assumption. 
Briefly, I start with a quantum coherent electromagnetic field state, obtain a classical electromagnetic field by going to a semiclassical limit, with the result that the quantum coherences within the quantum electromagnetic field state are being replaced with classical coherences in the classical electromagnetic wave. This wave, considered in all generality, is the particular "light beam" used by Einstein to define his causal structure. However, at this point, I construct a purification of this classical electromagnetic wave, in which I entangle it with suitable ancillas to reconstruct the classical correlations in the form of quantum correlations (entanglement) of the specific photon modes to the ancillas. At this point, control of the ancillas can be exercised, resulting in the creation of differences in phases and relative changes of coherence. These operations can be so tuned in order to obtain, upon tracing out the ancillas, a new classical electromagnetic wave, that however has a propagation speed larger than the propagation speed of the usual electromagnetic wave (light) in normal spacetime. Using this new effective speed, obtained strictly through alteration of the correlation structure at the level of the Uhlmann bundle, I will derive a resulting entropic force, related to the rate of change that can be induced to the effective light speed by altering the phases, a force that will play the role of an acceleration in normal spacetime, an effect of mass reduction, as well as access to an extended efficient light cone, beyond the usual light cone of classical special relativity. The overall effect is the creation of an effective Alcubierre-type bubble, in which the speed of light is quantum-engineered to exceed its standard value, enabling causal connections that would otherwise be impossible. This paper provides a concise practical calculation demonstrating that tools and techniques available in current or near-future tabletop quantum gravity experiments could suffice to increase the effective speed of light by approximately 3000 $km/s$. Crucially, this method does not rely on exotic matter or complex matter distributions, unlike the conventional Alcubierre framework. Additionally, the article proposes both an experimental verification of this phenomenon and a practical implementation using a tabletop setup.
\section{Uhlmann gauge theory} 
In order to construct what I presented in the introduction some auxiliary concepts are required. Among those, I will definitely need to work with quantum information geometry and emergent gauge fields. If a quantum system in a pure state $\ket{\psi(t)}$ evolves adiabatically and cyclically under a parametrised Hamiltonian $H(R(t))$ then the state picks up a Berry phase 
\begin{equation}
\ket{\psi(t)}\rightarrow e^{i\gamma}\ket{\psi(0)}
\end{equation}
where $\gamma=\oint_{l}\mathcal{A}(R)\cdot dR$ with $\mathcal{A}(R)=i\bra{\psi(R)}\nabla_{R}\ket{\psi(R)}$ the Berry connection and the curvature $\mathcal{F}=d\mathcal{A}$ gives rise to topological invariants (e.g. Chern number). The Berry connection behaves like a gauge field over parameter space. The parameter space represents external fields that are considered to be acting on the system. If we generalise this to mixed states, we obtain an additional phase structure. Pure states live in a Hilbert space modulo the phase 
\begin{equation}
\mathcal{H}_{phys}=\{projective \;\; Hilbert \;\; space\}
\end{equation}
but mixed states are described by density matrices $\rho$ which live in a different space, the space of positive semidefinite operators with trace 1. The Uhlmann phase is precisely the analogue of the Berry phase for the density matrix. It is in fact a geometric phase for density matrices. Given a family of density matrices $\rho(R)$, we define a purification of each $\rho$ namely 
\begin{equation}
\begin{array}{cc}
\rho=Tr_{E}\ket{\Psi}\bra{\Psi}, & \ket{\Psi}\in \mathcal{H}_{S}\otimes \mathcal{H}_{E}\\
\end{array}
\end{equation}
We then require for the purification to evolve in a parallel fashion in a certain metric induced connection over the state space (Uhlmann parallel transport). We define therefore a phase picked up under a cyclic evolution of $\rho(R)$. This defines the Uhlmann phase which generalises the Berry phase for mixed states. This phase depends on the geometry of the space of density matrices, and the fidelity metric, or the measure of quantum distance between states. The Uhlmann connection defines a notion of parallel transport over a base space (the parameter space), resulting in a non-Abelian gauge field defined over the manifold of mixed states. The Uhlmann curvature plays the role of the field strength. Therefore we have an emergent gauge field and we obtained an Uhlmann gauge bundle associated to the geometry of quantum state space. 
\begin{equation}
\begin{array}{cc}
\mathcal{A}^{Uhlmann}=\{Uhlmann \;\; connection\} \Rightarrow & \mathcal{F}^{Uhlmann}=d\mathcal{A}+\mathcal{A}\wedge \mathcal{A}
\end{array}
\end{equation}
In the usual approach, the Uhlmann gauge field is a geometric field. It arises when we consider the parallel transport over the space of density matrices parametrised by external parameters $\lambda$, a$\rho(\lambda)$ and which has a purification $\ket{\Psi}$. It defines a connection on a fibre bundle of purifications, it introduces a holonomy under cyclic paths $\rho(\lambda)\rightarrow \rho(\lambda+\delta \lambda)\rightarrow ...\rightarrow \rho(\lambda)$ and it defines a phase analogous to the Berry phase but for mixed states. However, such a connection doesn't appear from a Lagrangian or Hamiltonian for a field, in the form the gauge field appears for example in Maxwell's equations. It only reflects information theoretic geometry and it is constructed from state-dependent parallel transport, and not as usual from field equations. Therefore, it doesn't obviously have a dynamics of its own. We can however reformulate this framework so that we create a theory in which Uhlmann gauge appears as a dynamical field. I will do this here in order to obtain a better control over the relative phase transformations considered as Uhlmann gauge transformations on the ancillas, and to explore structures like quantum information space instantons that would help stabilise the modifies spacetime causal region, but it is not essential for the result. However, as a novelty, I will introduce it here and explain its usefulness in this and other cases. Let us consider a theory in which the Uhlmann gauge field $\mathcal{A}$ is a dynamical variable. I will write an action functional which will also include a kinetic term 
\begin{equation}
S[\mathcal{A}]=\int Tr(F\wedge *F)+\int \Psi^{\dagger}D_{\mu}\Psi
\end{equation}
The gauge field is considered to couple to the purification degrees of freedom and the path integration is over $\mathcal{A}$. This is now no longer a passive geometric structure, and in fact $\mathcal{A}$ has independent degrees of freedom, excitations, back-reaction, etc. 
In this way, we will transform information geometry into a field theory with emergent interactions. This is a new construction and therefore we have to carefully define the context of this theory. The "matter" fields here will not be actual matter fields, but instead they will be purifications, but on its own the Uhlmann connection will now be a dynamical field, fluctuating etc. 
If we introduce also the purifications (as "matter" fields) we obtain a theory of the form 
\begin{equation}
S[\mathcal{A},\Psi]=\int \Psi^{\dagger}D_{\lambda}\Psi + \frac{1}{4g^{2}}Tr(F_{\lambda\lambda'}F^{\lambda\lambda'})
\end{equation}
Where $\Psi(\lambda)$ are purifications, behaving like fields on the base manifold, and $\mathcal{A}$ is the Uhlmann connection now promoted to a field variable, where then the "covariant" derivative would be 
$D_{\lambda}=\partial_{\lambda}+\mathcal{A}_{\lambda}$ and the field strength (curvature) would become 
\begin{equation}
F_{\lambda \lambda'}=\partial_{\lambda}\mathcal{A}_{\lambda'} - \partial_{\lambda'} \mathcal{A}_{\lambda} +[\mathcal{A}_{\lambda}, \mathcal{A}_{\lambda'}]
\end{equation}
From this action we would get Euler-Lagrange equations for the gauge field $\mathcal{A}_{\lambda}$. These equations would determine how the Uhlmann connection evolves, and also how it mediates interactions between different purifications. We could also obtain propagating Uhlmann "gauge bosons" in the parameter space. On a deeper level, we could create a canonical structure by introducing commutation relations, quantised gauge models and coupling of matter (purifications or other quantum degrees of freedom) to the gauge field. In this way we could also study quantum informational topological sectors and even determine entanglement between phase transitions. The end result would be possibly a form of emergent interactions in open quantum systems. This would turn the information geometry into an effective physical theory and we would obtain emergent forces based on entanglement geometry and provide a field theoretic formulation of quantum information flow. The Uhlmann gauge field arises in the space of purification of mixed states. A density matrix is in general describing a mixed state (obtained for example from some partial state) while a purification would be formed in a larger Hilbert space formed from the system plus the ancilla, with the property that all purifications of $\rho$ are related by unitary rotations on the ancilla 
\begin{equation}
\ket{\Psi_{\rho}}\sim (I\otimes U)\ket{\Psi_{\rho}}
\end{equation}
and the Uhlmann connection defines the parallel transport over the manifold of purifications. The holonomy of this connection is the Uhlmann phase. The conclusion is that the Uhlmann gaug efield is a non-Abelian connection on a fibre (gauge) bundle over the manifold of density matrices. The "matter" fields are the purifications themselves which transform under the gauge group $U(n)$ (acting on the ancilla). They represent redundant descriptions of $\rho$ but they can evolve and interact. The Uhlmann gauge field would mediate interactions between them. The fact that the Uhlmann field has become dynamical can be understood in the sense that it can interact with different purification sectors and can create entanglement between such purifications. Therefore, the Uhlmann gauge field would encode and mediate entanglement and coherence relationships among mixed states and their purifications. This opens the possibility for an entanglement mediated interaction, an entropic force between reduced subsystems. Suppose we have two subsystems $A$ and $B$ each described by reduced density matrices $\rho_{A}$ and $\rho_{B}$ possibly derived from a larger purse state $\Psi_{AB}$. Each $\rho_A$ and $\rho_{B}$ has its own purification 
\begin{equation}
\begin{array}{c}
\ket{\Psi_{A}}\in \mathcal{H}_{A}\otimes \mathcal{H}_{E_{A}}\\
\ket{\Psi_{B}}\in \mathcal{H}_{B}\otimes \mathcal{H}_{E_{B}}\\
\end{array}
\end{equation}
Now suppose the system evolves along a path in parameter space and we define an Uhlmann connection for each purification bundle. Then the Uhlmann gauge fields can encode how these purifications entangle or disentangle, mediate geometric phase effects between sectors and define a new kind of coupling between entangled subsystems via a shared geometric informational structure. Moreover, a dynamical Uhlmann gauge field would model a force that is of purely informational origin acting between subsystems via their entanglement structure. The space of purifications can be seen as a $U(n)$ gauge fibre over the base space of mixed states $\rho$. When we say that two purifications interact it means that the coherence structure of one subsystem's mixed state can influence the purification (or coherence structure) of the other. This is not a traditional force between particles but instead a new kind of entanglement mediated coupling where the quantum geometry of purification space becomes a channel of influence. The interaction strength depends on the overlaps of purifications and the system would be sensitive to global consistency of entanglement structure. Therefore, if promoted to a dynamical field, the Uhlmann gauge field would mediate interactions between different purifications of mixed states, effectively encoding how entanglement, coherence, and information geometry evolve across parameter space. This would enable new kinds of interactions between subsystems of a quantum many body system governed not by spacetime locality but by the state-space geometry, which would imply a truly quantum informational force. We could for example form two reduced density matrices $\rho_{A}$ and $\rho_{B}$ derived from an entangled Bell state. We then purify those reduced states using square roots. A dynamical Uhlmann gauge field represented as a fluctuating field over a parameter space would be introduced, and we would obtain a toy model effective action in which we have a kinetic term (curvature penalty for the gauge field), an interaction term (entanglement mediated overlap between purifications) and a coupling constant for matter-gauge interactions. We could therefore compute an effective action in this context representing the energy of the system under this Uhlmann-like interaction. In this way we would start the construction of a field theory of quantum coherence. But what would physically mean for two different purifications of mixed states to interact via an Uhlmann-type gauge field? Given a mixed state $\rho_{A}$ a purification is a pure state in a larger Hilbert space, 
\begin{equation}
\ket{\Psi}\in \mathcal{H}_{A}\otimes \mathcal{H}_{E}
\end{equation}
such that $\rho_{A}=Tr_{E}\ket{\Psi}\bra{\Psi}$. However, this purification is not unique, all purifications are related by unitary transformations on the ancilla
\begin{equation}
\ket{\Psi'\rho}=(I\otimes U)\ket{\Psi\rho}
\end{equation}
Therefore the space of purifications is a $U(n)$ gauge fibre over the base space of mixed states $\rho$. If we construct two reduced density matrices from a shared entangled state and then purify them separately and let the Uhlmann gauge field be dynamical, then this gauge field can mediate an interaction between these two purifications. Physically, that means the coherence structure of one subsystem's mixed state can influence the purification or coherence structure of the other. This is not a usual interaction between particles, but a form of entanglement mediated coupling where the quantum geometry of purification space becomes a channel of influence, the interaction strength depends on overlaps of purification and the system is sensitive to global consistency of the entanglement structure. This explains why I called this quantum geometry previously a "channel of influence". This defines a new paradigm for quantum information interactions, one that is not based on particles or systems but instead on coherence. This interaction would not be mediated by fields in spacetime but by gauge fields over the state space. This would be able to model a series of phenomena, like quantum synchronisation, coherence transfer, decoherence spreading, or entanglement rigidity. In this sense the Uhlmann gauge field mediates "forces" between purifications i.e. between coherence sectors of mixed subsystems. 
The effective action in this toy model would look like 
\begin{equation}
S_{eff}=\frac{1}{2g^{2}}\int F^{2}+\int \Psi_{A}^{\dagger}\Psi_{B}
\end{equation}
In this action we have the fluctuations of the Uhlmann gauge field (an analogue to curvature), encoded by $F^{2}$, and a coherence based interaction term $\Psi_{A}^{\dagger}\Psi_{B}$. We would encode highly aligned states $\Psi_{A}$ and $\Psi_{B}$ as having a low action, and therefore a coherent global alignment, and if they are misaligned, it would imply a higher energy cost. For two purifications of different subsystems to interact via the Uhlmann gauge field it would mean that their coherence structures are coupled through a shared entanglement geometry mediated by a non-Abelian connection on the purification bundle. This "interaction" would not be in spacetime but instead it would a state-space interaction, which would amount to a new type of quantum informational force that would govern entanglement space consistency and coherence propagation. 
In a pure state $\ket{\psi}$ coherence refers to quantum superposition between basis states. In a mixed state $\rho$ coherence has a different interpretation. The off-diagonal elements in a density matrix defined over a basis of classically distinguishable states represent in this particular basis, quantum coherence between classical alternatives. Therefore a coherence sector is a block of the Hilbert space (or a sector of the density matrix) where such off-diagonal quantum correlations are supported. Let us therefore consider our larger quantum system $A+B$ in a pure entangled state. If we now trace over B we obtain the subsystem A as being in a mixed state $\rho_{A}$. This $\rho_{A}$ might be diagonal and hence fully coherent in a given basis, and in that case we would have no coherence, or have off diagonal elements, case in which we would have residual quantum coherence from the entanglement with $B$. In the previous example I obtained $\rho_{A}$ and $\rho_{B}$ from tracing over an entangled Bell pair. The purifications $\sqrt{\rho_{A}}$ and $\sqrt{\rho_{B}}$ encode the residual structure of coherence in their respective subsystems. These purifications live in sectors of extended Hilbert space and the Uhlmann field tracks how their relative coherence structure evolves. When we say that an Uhlmann field mediates interactions between coherence sectors of mixed subsystems, this means that each purification represents a coherent embedding of a mixed state in a larger space and the way one purification aligns or misaligns with another affects their relative phase structure. The overlap is calculated by $Tr(\sqrt{\rho_{A}}\sqrt{\rho_{B}})$ measures how coherent their relation is and the Uhlmann gauge field geometrizes this relation. This is important in a series of physical cases where subsystems with residua coherence influence each other's coherence or where deocherence in one region can affect the phase structure of another and coherence transfer is mediated by an entanglement structure and not spatial interaction. Indeed, a few physical situations where this actually happens are open quantum systems, quantum thermodynamics, quantum information flow or topological quantum matter at finite temperature. 
In general, for any mixed state $\rho$, the purifications of $\rho$ form a fibre bundle where the base space is the space of density matrices and the fibres are all purifications over $\rho$ related by $\ket{\Psi_{\rho}}\sim (I\otimes U)\ket{\Psi_{\rho}}$. This bundle is an Uhlmann bundle and its connection defined the associated parallel transport along paths in the base space. Therefore the coherence structure of a mixed state isn't just what is defined in $\rho$ but also the information encoded in how its purifications vary and hence in their geometry and holonomy. We therfore introduce the gauge field $A_{\lambda}$ over the parameter $\lambda$ and we add the coupling term as defined above, an effective overlap of purifications
\begin{equation}
Tr(\sqrt{\rho_{A}}\sqrt{\rho_{B}})
\end{equation}
This would imply that the degree of purification alignment across subsystem matters and it can contribute to the effective energy. This would model how entanglement between parts of a system constraints their internal structure and how global coherence emerges from geometrically structured interactions between subsystems' purifications. 
The Uhlmann gauge field describes parallel transport of purifications and therefore once promoted to a dynamical field it would mediate the way in which the geometry of one purification affects the other, and this defines a form of phase rigidity or entanglement stiffness across subsystems. A change in the purification of $\rho_{A}$ would cause a curvature that affects $\rho_{B}$. Therefore, the base space is made out of the usual density matrices, the fibres are the purifications, the metric is the Uhlmann (Bures) metric, the connection would be an Uhlmann connection, and the curvature is represented by the Uhlmann field strength. This structure encodes how coherence evolves along paths in parameter space, allows the definition of Uhlmann holonomies and can be interpreted as a non-Abelian gauge field over information space. If for the sake of an example we would have two superfluid regions with independent phases, if we tried to connect them, their relative phase mismatch would create an energy cost and the gauge field (namely phase difference) would mediate the stiffness between them. In our case, the purifications are like phase carrying states. The Uhlmann connection would measure their relative twist, and a dynamical Uhlmann field would impose a penalty for misalignment (an interaction). This generalisation can bring us even further. What would for example the a charge and an equivalent of a "coulomb field" in this situation? In general the charge is a source of the gauge field, and it enters Gauss' law which is basically a constraint, $\nabla\cdot\vec{E}=\rho_{charge}$. Such a charge would generate a gauge field, say a Coulomb field. The analogue for the Uhlmann gauge model would be that the matter fields are purifications of density matrices, the interaction terms appears as a trace of a product of two purifications, and this coupling plays the role of a gauge interaction. The gauge charge for a density matrix would in this case be a measure of how much its purification geometry couples to the Uhlmann gauge field. The alignment of purifications across subsystems defines a charge like property. If the purification of $\rho_{A}$ is twisted relative to $\rho_{B}$, it sources a holonomy. The charge is a measure of decoherence induced phase difference or some form of entanglement misalignment. 
We could therefore define a local Uhlmann gauge charge density by analogy as 
\begin{equation}
Q(\lambda)=\Psi^{\dagger}(\lambda)\mathcal{G}(\lambda)\Psi(\lambda)
\end{equation}
where $\mathcal{G}$ is the generator of gauge transformations, and $\Psi(\lambda)$ is the purification. This defines essentially how strongly the gauge field must twist to maintain coherence, therefore an effective charge density. The analogue of the Coulomb field is the non-trivial Uhlmann connection $\mathcal{A}_{\lambda}$ generated by a twisted purification. If we change the purification of a density matrix (for example by rotating the phase structure) we generate a non-zero Uhlmann curvature $\mathcal{F}_{\lambda\lambda'}$ and this is expected to affect the parallel transport of neighbouring purifications. Therefore the field produced by a coherence-charged purification is the Uhlmann gauge field $\mathcal{A}$ whose curvature $\mathcal{F}$ encodes the field strength and contains information analogous to the QED $\vec{E}$. It is worth mentioning for the sake of consistency that the coupling side appears with a negative sign, therefore if two purifications share the same relative phase structure then 
\begin{equation}
Tr(\sqrt{\rho_{A}}\cdot \sqrt{\rho_{B}})=1
\end{equation}
but the term appears in the action with a negative sign. The maximal Uhlmann fidelity implies they are parallel in the Uhlmann bundle and then they do not provoke an energy "penalty" as opposed to them being misaligned. If on the other hand $\rho_{A}=\rho_{B}$ but we apply different unitary transformations on their ancilla subsystems, the purifications may differ
\begin{equation}
\begin{array}{cc}
\ket{\Psi_{A}}=\sum_{k}\sqrt{p_{k}}\ket{k}\otimes \ket{e_{k}}, & \ket{\Psi_{B}}=\sum_{k}\sqrt{p_{k}}\otimes U\ket{e_{k}}
\end{array}
\end{equation}
We notice that the mixed state are the same but the purifications are different, therefore their overlap is reduced and they are misaligned. 

\section{The classical limit of quantum electrodynamics, what is light?}
One of the most important achievements of 20th century physics was the construction of a quantised theory of electromagnetism. Not only had a theory of quantum electrodynamics be constructed but also, a perturbative solution had to be implemented in order to obtain meaningful solutions. 
However, once we constructed such a quantum theory, another problem emerged, namely that of constructing back a classical (or semiclassical) structure that would be equivalent to what has previously been called "light". This problem is essentially not trivial due to the various ways in which a quantum to classical limit can be performed. Therefore we can consider that we start with a gauge field in QED and we want to take some semiclassical limit in order to obtain a (semi)classical electromagnetic wave. A brief discussion of the context as is understood today is presented in the following. In QED the electromagnetic field is described as a quantised gauge field represented by the vector potential operator $\hat{A}_{\mu}(x)$. Its behaviour is captured by field operators satisfying commutation relations and is described by a quantum field theory with the Lagrangian 
\begin{equation}
\begin{array}{cc}
\mathcal{L}_{QED}=\bar{\psi}(i\gamma^{\mu}D_{\mu}-m)\psi-\frac{1}{4}F_{\mu\nu}F^{\mu\nu}, & F_{\mu\nu}=\partial_{\mu}A_{\nu}-\partial_{\nu}A_{\mu}
\end{array}
\end{equation}
where all the terms are known and the gauge covariant derivative is $D_{\mu}=\partial_{\mu}+ieA_{\mu}(x)$. The fields in QED are operator valued distributions subject to the quantisation conditions written in the form of canonical commutation relations for gauge fields or canonical anti-commutation relations for fermions. 
The semiclassical limit involves considering quantum fluctuations as very small and describing the field primarily by its classical expectation values, augmented by small quantum corrections. This limit is often relevant for high intensity coherent states such as laser beams where the photon occupation number per mode is very large and quantum fluctuations become negligible in relative terms. 
Therefore it seems like a suitable quantum state for taking the semiclassical limit is a coherent state $\ket{\alpha}$ which we write as 
\begin{equation}
\hat{a}_{k,\lambda}=\alpha_{k,\lambda}\ket{\alpha}
\end{equation}
coherent states are therefore seen from a technical point of view as eigenstates of annihilation operators $\hat{a}_{k,\lambda}$ with eigenvalues $\alpha_{k,\lambda}$. I will show later on how this technical approach is actually an approximation, but at this point I am just presenting a quick introduction into the subject. Much will change as we continue advancing in the subject. These states as defined up to now, minimise uncertainty, representing states that are in a sense closest to classical fields. Coherent states generally describe laser radiation, classical radio waves, etc. 
Given such a quantum gauge field operator $\hat{A}_{\mu}(x)$, define its expectation value in a coherent state
\begin{equation}
A_{\mu}^{cl}(x)=\bra{\alpha}\hat{A}_{\mu}(x)\ket{\alpha}
\end{equation}
This expectation value is the classical (or semiclassical) gauge field associated with the quantum coherent state. Small fluctuations around this classical solution correspond to quantum corrections. 
A semiclassical electromagnetic wave is precisely the expectation value of the quantum field in a coherent state. Such a wave has a well defined amplitude and phase, closely mimicking classical electromagnetic radiation. Its residual quantum fluctuations can, in principle, affect phenomena sensitive to quantum correlations (for example photon counting experiments, squeezing phenomena, quantum optics effects). 
A classical electromagnetic wave is described entirely by classical Maxwell equations, without explicit reference to quantum operators. Therefore it is represented purely by c-number fields $A_{\mu}(x)$ which are solutions to Maxwell's classical equations
\begin{equation}
\begin{array}{cc}
\partial_{\mu}F^{\mu\nu}(x)=0, & with \;\; F_{\mu\nu}(x)=\partial_{\mu}A_{\nu}(x)-\partial_{\nu}A_{\mu}(x)
\end{array}
\end{equation}
A classical wave is fully deterministic and has no quantum fluctuations, therefore is expected to exhibit exact coherence properties. We know that there are multiple semiclassical limits and different ways to go from quantum to classical physics, sometimes we even find out that a direct classical limit doesn't even exist. 
We can for example in this introductory approach consider that the connection between the semiclassical and classical description arises through the large-photon-number limit, and through the diminishing of the relative quantum fluctuations. 
Let us consider a coherent state with large photon occupation number per mode
\begin{equation}
|\alpha_{k,\lambda}|^{2}\gg 1
\end{equation}
Quantum fluctuations scale like $\sim \frac{1}{\sqrt{|\alpha|^{2}}}$. As photon numbers grow, relative fluctuations diminish
\begin{equation}
\begin{array}{cc}
\frac{fluctuations}{signal}\sim\frac{1}{|\alpha|}\rightarrow 0 & as |\alpha|\rightarrow \infty
\end{array}
\end{equation}
In this limit the expectation value for the gauge field becomes effectively indistinguishable from a purely classical solution of Maxwell's equations and quantum fluctuations vanish. This amounts to quantum expectation values turning smoothly into classical fields and quantum commutators becoming negligible at the macroscopic level. 
Let us analyse the discussion up to this point in more details in order to clarify some aspects that will be important later on. Usually, a coherent state in quantum optics and in quantum field theory is a special quantum state that represents the quantum analogue of a classical electromagnetic wave. It can be defined as an eigenstate of the annihilation operator $\hat{a}$ given by $\hat{a}\ket{\alpha}=\alpha\ket{\alpha}$, $\alpha\in\mathbb{C}$. The coherent state has a definite amplitude and phase. It also minimises the product of uncertainties in the canonical conjugate variables (say electric and magnetic fields, etc.). The coherent state is also stable under field evolution, hence under a linear Hamiltonian they may only pick up a phase factor in their eigenvalue. The photon number distribution of such a coherent state is Poissonian, namely the probability $P(n)$ of finding exactly $n$ photons in a coherent state follows a Poisson distribution 
\begin{equation}
P(n)=e^{-|\alpha|^{2}}\frac{|\alpha|^{2n}}{n!}
\end{equation}
In general coherent states form a continuous, non-orthogonal, overcomplete set making them useful for representing fields in quantum optics. Coherent states are generally useful because they have a well defined classical limit as their amplitude (the photon number) becomes very large. Quantum fluctuations around their mean values become relatively negligible as the intensity increases. It is interesting to notice however some technical aspects often overlooked. In general, if one applies a photon destruction operator on a Fock state one would obtain 
\begin{equation}
\hat{a}\ket{n}=\sqrt{n}\ket{n-1}
\end{equation}
thus a coherent state cannot simply be a fixed number state. If it were so, it wouldn't satisfy the technical condition we demanded from it, namely to conserve its shape after applying a photon destruction operator. In fact the actual coherent state is represented as a superposition of an infinite number of many photon states. 
\begin{equation}
\ket{\alpha}=e^{-|\alpha|^{2}}{2}\sum_{n=0}^{\infty}\frac{\alpha^{n}}{\sqrt{n!}}\ket{n}
\end{equation}
The coherent state is therefore a sum over all possible photon number states weighted by a Poisson distribution. It is never a state with a fixed photon number, but one with an infinite photon number having certain probabilities to observe each of those photon numbers. 
If we apply on this state a photon annihilation operator we obtain 
\begin{equation}
\hat{a}\ket{\alpha}=e^{\frac{|\alpha|^{2}}{2}}\sum_{n=0}^{\infty}\frac{\alpha^{n}}{\sqrt{n!}}\hat{a}\ket{n}=e^{\frac{|\alpha|^{2}}{2}}\sum_{n=0}^{\infty}\frac{\alpha^{n}}{\sqrt{n!}}\sqrt{n}\ket{n-1}
\end{equation}
and by changing the summation index $m=n-1$ we obtain 
\begin{equation}
e^{\frac{|\alpha|^{2}}{2}}\sum_{n=0}^{\infty}\frac{\alpha^{m-1}}{\sqrt{(m+1)!}}\sqrt{m+1}\ket{m}
\end{equation}
after simplifying the factorial factor we simply obtain 
\begin{equation}
\alpha e^{\frac{|\alpha|^{2}}{2}}\sum_{n=0}^{\infty}\frac{\alpha^{m}}{\sqrt{m!}}\ket{m}=\alpha\ket{\alpha}
\end{equation}
The key observation here is that the state has no definite photon number. It instead contains many photon number components with the distribution centred around an average photon number $\bar{n}=|\alpha|^{2}$. Removing only one photon simply shifts the distribution slightly down but due to the special structure of the coherent state, this shifted distribution is identical to the original distribution. The coherent state is exactly constructed such that adding or removing a photon doesn't change its overall shape, just scales it. Such a special property only emerges from the particular Poissonian superposition of photon number states. 
However, when constructing such a coherent state, we rely on the sum going to infinity, basically taking into account Fock states with an infinite number of particles in the limit. In the case of a finite (and hence more realistic) sum, we wouldn't strictly have such a coherent state anymore. The existence of an infinite superposition of photon number (Fock) states is therefore essential. If the sum were to be truncated, then we wouldn't obtain an exact eigenstate anymore, instead only an approximate one. 
Therefore a realistic superposition of Fock states would not be technically coherent. This aspect brings us to some limitations of quantum field theory. In principle, a pure QED alone, without gravity or other fields, would have the photon as a massless, non-interacting gauge boson. This would imply that photons do not experience self-interaction directly (and therefore they are not interacting at tree level in QED, although we know photons self-interact via square-diagrams at higher orders). As a result, in pure QED when we ignore gravity, there is no intrinsic upper bound to how many photons can be introduced in a certain region of spacetime. However, this idealised scenario neglects a series of important physical realities, in particular gravitational and quantum gravitational effects. 
Once gravity is taken into consideration, even just classical gravity described by Einstein's field equations, we notice that photons carry energy proportional to their frequency and thus contribute to the stress energy tensor, acting as a source of gravity. If we attempt to pack too many photons and hence too much energy density into a limited spatial region, the region will eventually become gravitationally unstable and will collapse into a black hole. The presence of gravity will therefore impose an upper bound on the energy density and thus on photon number densities within a given finite region of spacetime. 
A physically meaningful bound arises from the Bekenstein bound which states that the amount of information and implicitly of energy one can contain within a region of radius R is bounded by black hole physics. The maximum energy that can be placed inside a sphere of radius R without it becoming a black hole is approximately the mass energy associated with a black hole of Schwarzschild radius R, namely 
\begin{equation}
E_{max}\sim \frac{Rc^{4}}{2G}
\end{equation}
where $G$ is Newton's gravitational constant, and $c$ is the speed of light. 
An approximate estimation of the maximum photon number goes as follows. Given a finite region of space of radius $R$ the maximum total energy allowed before gravitational collapse into a black hole is approximately $E_{max}=\frac{c^{4}}{2G}R$. If we assume photons of frequency $\nu$, each photon having the energy $E_{\gamma}=h\nu$. Thus the maximum photon number $N_{max}$ in a volume of radius $R$ (assuming monochromatic photons), is 
\begin{equation}
N_{max}=\frac{E_{max}}{h\nu}=\frac{c^{4}R}{2Gh\nu}
\end{equation}
If we consider the information content, the Bekenstein Hawking entropy bound explicitly limits the maximum number of quantum states, namely any finite region of space has a maximum entropy and therefore a maximum quantum state occupation number, bounded by the surface area of the region 
\begin{equation}
S_{BH}\leq \frac{k_{B}c^{3}}{\hbar G}\frac{A}{4}
\end{equation}
This also translates into limits on the quantum states accessible for photons within the region. Therefore quantum field theory with no cutoff predicts infinitely many photon modes, as frequency grows without bound. But physically there are natural cutoffs, namely a UV cutoff given by Planck frequency $\nu_{Planck}\sim 10^{43}Hz$ which limits photon energies, and the IR cutoff given by the finite size of the region which sets the lowest frequency mode available. A brief practical estimation would be, considering the radius of a spherical region to be of 1 meter
\begin{equation}
E_{max}\sim \frac{c^{4}R}{2G}\sim \frac{(3\times 10^{8} m/s)^{4}\times 1 m}{2\times 6.67\times 10^{-11} Nm^{2}/kg^{2}}\sim 10^{43}\;\; Joules
\end{equation}
and for visible photons $(\nu\sim 5\times 10^{14} Hz)$, $E_{\gamma}\sim 3\times 10^{-19}$ joules, the maximum photon number is roughly 
\begin{equation}
N_{max}\sim \frac{10^{43}}{3\times 10^{-19}}\sim 10^{62} \;\; photons
\end{equation}
However, in this simple approximation we calculated the number of photons of a very specific frequency. In reality the maximum number of photons will be different if we considered a Poissonian distribution as it is required for a coherent state. A coherent state would not be monochromatic in the strict photon number sense, but instead it will follow a Poissonian distribution with a mean photon number $\bar{n}=|\alpha|^{2}$. Such a coherent state $\ket{\alpha}$ has photon number probabilities given by 
\begin{equation}
P(n)=e^{-\bar{n}}\frac{\bar{n}^{n}}{n!}
\end{equation}
The mean photon number is $\bar{n}$ and the fluctuations around $\bar{n}$ are of the order $\sqrt{\bar{n}}$. The coherent state is not fixed in photon number but instead it is centred around $\bar{n}$ with fluctuations small relative to $\bar{n}$ when $\bar{n}$ is large. Again, to avoid gravitational collapse into a black hole, the region of radius R must not contain energy greater than the Schwarzchild bound. Let us again consider only a single mode approximation for now. A coherent state is usually defined for a single mode of electromagnetic radiation with frequency $\nu$, and thus photon energy $h\nu$. The mean photon number in that single mode would be limited by the gravitational energy bound as before, $\bar{n}_{max}=\frac{E_{max}}{h\nu}$. This is essentially identical with the single frequency calculation above because a coherent state defined over a single mode frequency will not differ from the earlier calculation. The maximum mean photon number is the same. But in practice, coherent states have multiple modes and frequency bandwidths. The coherent states usually occupy multiple modes (in frequency) and a multi-mode coherent state is simply a product of single mode coherent states 
\begin{equation}
\ket{\{\alpha_{k}\}}=\otimes_{k}\ket{\alpha_{k}}
\end{equation}
each mode $k$ having mean photons $|\alpha_{k}|^{2}$. 
Thus the total mean photon number would be 
\begin{equation}
\bar{N}_{total}=\sum_{k}|\alpha_{k}|^{2}
\end{equation}
subject to the total energy constraint 
\begin{equation}
\sum_{k}|\alpha_{k}|^{2}h\nu_{k}\leq \frac{c^{4}R}{2G}
\end{equation}
To maximise the total number of photons we should in principle distribute the photons preferably into lower-energy, long wavelength modes. The finite spatial volume of radius R introduces a natural IR cutoff, the wavelength $\lambda_{max}\cong 2R$ and hence the minimum frequency $\nu_{min}\cong \frac{c}{2R}$. which brings us to the photon energy at its lowest frequency mode $h\nu_{min}\cong \frac{hc}{2R}$. Then the maximal photon number (mean) allowed in this lowest frequency mode is 
\begin{equation}
\bar{N}_{max}=\frac{E_{max}}{h\nu_{min}}=\frac{c^{3}R^{2}}{Gh}
\end{equation}
This is the maximal fundamental limit of the photon number, achievable if you distribute photons optimally into the lowest energy mode permitted by the size of the region. Numerically, this amounts to (if we take again R=1 meter) 
\begin{equation}
\bar{N}_{max}=\frac{(3\times 10^{8})^{3}\times (1)^{2}}{(6.67\times 10^{-11})\times (6.6\times 10^{-34})}
\end{equation}
This gives approximately 
\begin{equation}
\bar{N}_{max}\sim 10^{77} \;\; photons
\end{equation}
This calculation gives only the mean number. A coherent state has fluctuations of the order $\sqrt{\bar{N}}$ which is negligible compared to the mean number itself. It is clear that the maximal photon number dramatically increases if we shift the photons towards lower frequencies. The fundamental limit however seems to depend on the frequency distribution. To maximise the photon number, you place photons at the lowest possible frequency allowed by the finite volume. 
However, once we pushed all the photons in the limit to the same frequency, this state is not strictly coherent in the technical sense. How can we possibly have coherence if all the photons are at the lowest frequency? Isn't that just a Fock state? Indeed, a coherent state is always a superposition of photon number states at one given mode (frequency). It doesn't appear usually as a superposition of frequencies. Coherence does not come from mixing different frequencies but from mixing different photon number states at one frequency. Therefore placing all photons in the lowest possible allowed frequency mode is consistent with coherence. We still have a Poisson distribution of photon numbers at that single frequency. It is still not a Fock state (which has a definite number of photons), it is coherent with indefinite photon numbers (a Poissonian distribution) but at only one single lowest energy mode. The fact that we took an extreme form of Poisson distribution in our approximation should not in principle change the assumption we made about the coherent state, namely that it is a superposition of Fock states. The distribution would simply degenerate into one that represents only one frequency mode. Photon number states, namely Fock states, have a definite number of photons but no definite phase coherence. Coherent states at a given frequency have definite phase coherence and indefinite photon numbers, they are single mode states in frequency but multi-number superpositions in photon number. Multiple frequencies would form a multi-mode coherent state but a single mode coherent states are still perfectly valid coherent states. However, what happens here is that we obtain a natural truncation imposed by physics, and hence a coherent state becomes only technically approximately coherent. Maximising photon number by placing photons in the lowest allowed frequency mode doesn't break coherence directly (frequency coherence is independent of photon number coherence) but a finite photon number truncation explicitly does break down exact coherence. 
This seems to be somewhat out of context in this paper, but all the details presented above will be needed as the article progresses. 
Because of this future utility, let us go further and describe a multi-mode coherent state more carefully. 
Such a state would be a direct product of single mode coherent states, each mode labeled by the frequency (or momentum) index $k$
\begin{equation}
\ket{\{\alpha_{k}\}}=\bigotimes_{k}\ket{\alpha_{k}}=\bigotimes_{k}e^{-\frac{|\alpha|^{2}}{2}}\sum_{n_{k}=0}^{\infty}\frac{\alpha_{k}^{n_{k}}}{\sqrt{n_{k}!}}\ket{n_{k}}
\end{equation}
with $\hat{a}_{k}\ket{\alpha_{k}}=\alpha_{k}\ket{\alpha_{k}}$. 
Such a multi-mode state would have independent Poissonian distributions in each mode $k$, each with a mean photon number in mode $k$, $\bar{n}_{k}=|\alpha_{k}|^{2}$, and a total mean number of photons $\bar{N}_{tot}=\sum_{k}|\alpha_{k}|^{2}$. Each frequency mode would have a definite, independent phase and amplitude, the relative phases between modes are also definite, due to the product structure, creating a coherent superposition across frequencies. We therefore have many frequency modes, each frequency being independently coherent. The exact eigenstate of each mode's annihilation operator satisfies the technical coherence constraint. A multi-mode coherent state is not an eigenstate of a single annihilation operator, rather, it is simultaneously an eigenstate of each mode's annihilation operator. This result would represent classical-like electromagnetic fields having multiple frequencies simultaneously. This is our typical "light pulse", a coherent superposition of different frequency modes. Such a multi-mode state is not less coherent, of course. Multi-mode states are coherent just as single mode coherent states. They are simply coherent states of multiple independent modes simultaneously. Coherence is therefore here defined by a definite amplitude and phase in each mode, not photon number. The photon number uncertainty is present in both single-mode and multi-mode coherent states, what matters is phase coherence which remains perfectly well defined. 
When we say that a state has phase coherence, it means that the quantum field described by that state has a well defined phase relationship. In the case of a single mode state, the quantum state has a definite relative phase between photon-number states. A coherent state $\ket{\alpha}$ explicitly has a well defined global phase and amplitude
\begin{equation}
\ket{\alpha}=e^{-|\alpha|^{2}/2}\sum_{n=0}^{\infty}\frac{\alpha^{n}}{\sqrt{n!}}\ket{n}=e^{-|\alpha|^{2}}/2\sum_{n=0}^{\infty}\frac{|\alpha|^{n}e^{i n\phi}}{\sqrt{n!}}\ket{n}
\end{equation}
where $\alpha=|\alpha| e^{i\phi}$. Each photon number component has a fixed relative phase $e^{i n \phi}$ therefore the state has phase coherence. For a multi-mode phase coherence, we have a multi-mode state
\begin{equation}
\ket{\{\alpha_{k}\}}=\bigotimes_{k}\ket{\alpha_{k}}
\end{equation}
Each mode individually has a definite phase. Additionally one has a clear relative phase relationship between the different modes, giving rise to classical like interference phenomena (pulses, wave packets, frequency combs). We can now see that the eigenstate property (the exact technical coherence criterion) guarantees coherence but is not the only way to achieve coherence. The much more general physical and operational criterion for coherence is phase coherence, which refers to the quantum state's property of having a clearly defined phase relationship between photon number components leading to classical like electromagnetic wave behaviour. This coherence is what allows for classical interference, superposition, and other wave-like behaviours in quantum optics. Phase coherence is the physically measurable property via interferometry or homodyne detection, whereas being an exact eigenstate of an annihilation operator is a more abstract mathematical, and often impractical condition. Being an exact eigenstate of the annihilation operator is mathematically elegant and uniquely identifies coherent states, however, in reality, many physically relevant states (such as truncated coherent states) do not exactly satisfy the eigenstate property but still behave with high coherence practically. Therefore phase coherence is a more general and practically relevant criterion. In fact a state can have phase coherence without being an exact annihilation operator eigenstate. Truncated coherent states described above are one example, but squeezed states also have a clearly defined phase relationship, but they are not eigenstates of $\hat{a}$, yet they are considered coherent-like states (with modified uncertainties). Operationally, in quantum optics, phase coherence means experimentally measurable interference properties, like homodyne detection, namely if we mix the quantum field with a local oscillator (reference field) the resulting interference pattern clearly reveals a well defined, stable relative phase. Also in interference experiments a coherent quantum state interferes just like a classical electromagnetic wave creating stable predictable interference fringes. Coherent states maximise coherence functions $g^{(1)}(\tau)$, closely matching classical coherence theory. Therefore coherence physically means stability and definiteness of wave-like interference properties. In multi-mode coherent states we have multiple independent frequency modes, each mode individually being phase coherent, but crucially we also have clearly defined relative phases between different frequency modes. This means we can form complicated classical-like optical pulses and wave packets directly resembling classical electromagnetic signals. This is why phase coherence matters especially in multi-mode coherent states: it ensures that we can synthesise classical-like electromagnetic waveforms from quantum fields. 
It is important to notice that the coherent state is in a sense a bridge between two types of coherence: the classical coherence and the quantum coherence. Quantum mechanically, a coherent state is a delicate superposition of photon-number states with precise quantum phases. This is what we call quantum coherence. Classically, this same coherent state corresponds precisely to a classical electromagnetic wave with well defined amplitudes and phases. Thus it also possesses classical coherence. The coherent state is the unique state that simultaneously possesses both forms of coherence clearly and strongly. The process of decoherence is that process in which quantum coherence is lost but classical coherence is maintained. When a coherent state interacts with a realistic environment (e.g. scattering with particles, coupling to electromagnetic environment etc. ) quantum coherence (superposition of photon-number states) rapidly becomes entangled with the environment and effectively washes out. The density matrix becomes diagonal in the photon number basis losing quantum phase information. Decoherence thus removes the delicate quantum superposition structure in photon number states, destroying quantum coherence. However, classical coherence, the macroscopical wave-like behaviour of the field, usually remains intact or is minimally affected. This is because classical coherence arises from average expectation values and classical-like phase stability, not the delicate quantum superpositions between different photon number states directly. Decoherence typically transforms the quantum coherent superposition (pure state) into a mixture of coherent states with different phases or amplitudes, but each coherent state individually remains classical and coherent in the optical sense. While decoherence destroys quantum coherence, it typically leaves the system in states still very close to classically coherent states or mixtures thereof. Such mixtures behave classically. 
Consider initially the coherent state density matrix that is pure and in a photon number basis looks like 
\begin{equation}
\rho_{coh}=\ket{\alpha}\bra{\alpha}=\sum_{n,m}\frac{\alpha^{n}(\alpha^{*})^{m}}{\sqrt{n!m!}}e^{-|\alpha|^{2}}\ket{n}\bra{m}
\end{equation}
This state has clear off-diagonal terms (quantum coherence), showing explicit quantum superpositions. After decoherence with the environment (photon scattering or absorption/re-emission) the density matrix becomes approximately diagonal 
\begin{equation}
\rho_{decoh}\sim \sum_{n}P(n)\ket{n}\bra{n}
\end{equation}
The quantum coherence disappears (the off diagonal terms) and now we have a mixed state with photon-number probabilities $P(n)\sim e^{-|\alpha|^{2}}\frac{|\alpha|^{2n}}{n!}$. This mixed state is classical-like in the photon number sense, as it has no explicit quantum coherence, yet, each classical wave (coherent state) is hidden by this diagonalization. We lost the explicit coherent quantum superposition structure. 

Let us consider the coherent single frequency state again 
\begin{equation}
\ket{\alpha}=e^{-\frac{|\alpha|^{2}}{2}}\sum_{n=0}^{\infty}\frac{|\alpha|^{n}e^{i n \phi}}{\sqrt{n!}}\ket{n}
\end{equation}
with $\alpha=|\alpha|e^{i\phi}$. Each photon number state $\ket{n}$ has a well defined relative phase $e^{i n \phi}$. However, by well defined here we do not mean the same absolute global phase for all the components. Instead, we mean the phase differences between successive photon number states are fixed and constant. For example when going from $\ket{n}$ to $\ket{n+1}$ we will have to multiply by the factor $e^{i\phi}$. This constant incremental phase factor defines a stable, coherent phase relationship. This well defined phase means basically just a constant relative phase increment between successive photon-number components in a single mode, the mode that is formed via quantum mechanical superpositions of photon number states. The relative incremental phase between different photon number states is always the same, and it is independent of the photon number itself. However, the coherent state evolves in time, acquiring a phase evolution $e^{-i\omega t}$ but at any fixed instants, the relative phase among photon-number states is fixed. The concept of phase coherence relies on the photon-number basis and the coherent state representation. Measuring in photon number alone does not show coherence directly, instead coherence becomes evident in interference measurements or homodyne detection. Within a multi-mode coherent state 
\begin{equation}
\ket{\{\alpha_{k}\}}=\bigotimes_{k}\ket{\alpha_{k}}=\bigotimes_{k}(e^{-\frac{|\alpha_{k}|^{2}}{2}}\sum_{n_{k}=0}^{\infty}\frac{|\alpha_{k}|^{n_{k}}e^{i n_{k}\phi_{k}}}{\sqrt{n_{k}!}}\ket{n_{k}})
\end{equation}
each frequency mode $k$ has its own well defined phase $\phi_{k}$. Within each mode, photon number states have a fixed incremental phase difference $\phi_{k}$. Multi-mode coherence also means the relative phases between different frequency modes are well defined and stable. Each mode has a fixed phase $\phi_{k}$ and these phases don't fluctuate randomly. This allows classical interference between different frequencies such as forming pulses, wave packets or frequency combs. Each mode has its own fixed phase, and the relative phase differences between frequency modes $(\phi_{k}-\phi_{k'})$ remains stable. If the relative phase between modes would be random or fluctuating we would loose coherence between frequencies and no stable interference pattern would form. In a standard multi-mode coherent state, independence means factorization of the total quantum state into separate single-mode coherent states
\begin{equation}
\ket{\{\alpha_{k}\}}=\ket{\alpha_{1}}\otimes \ket{\alpha_{2}}\otimes... 
\end{equation}
There is no quantum correlation (no entanglement, no quantum correlation in photon number fluctuations) between different frequency modes. Each frequency mode can be manipulated and measured separately without affecting the other modes. Photon number fluctuations in one mode don't affect photon number fluctuations in another. 
However, classical correlation is present in standard multi-mode coherent states because we have definite fixed relative phases between modes. However, the state is still factorised quantum mechanically, there is therefore no quantum entanglement. 
It is possible still to produce quantum states of light where frequency modes are quantum mechanically entangled. Such states are not simple tensor products of coherent states, instead they are usually created by non-linear optical processes, like spontaneous parametric down-conversion, four-wave mixing or squeezing. For example a two mode squeezed vacuum state 
\begin{equation}
\ket{TMSV}\sim \sum_{n}c_{n}\ket{n}_{\omega_{1}}\ket{n}_{\omega_{2}}
\end{equation}
this state clearly has photon number correlations between modes. 
We could also have frequency bin entangled states, used in quantum communication and quantum computing 
\begin{equation}
\ket{\psi}=\frac{\ket{1}_{\omega_{1}}\ket{0}_{\omega_{2}}+\ket{0}_{\omega_{1}}\ket{1}_{\omega_{2}}}{\sqrt{2}}
\end{equation}
In such states, the photon number statistics of one mode depends strongly on the measurement outcome in another mode, indicating genuine quantum correlation (entanglement). Thus multi-mode states can have quantum correlations between frequency modes but such states are fundamentally different from the simple product of coherent states considered initially. 
A special type of measurement is the measurement in photon number alone. This measurement is performed using a photon number (Fock) basis and is typically done using a photon counting detector. The question asked by such a detector is "how many photons are present in a given mode?" and the measurement collapses the quantum state onto a definite photon number state $\ket{n}$. Such a photon number measurement discards all phase information. Photon number states do not carry any intrinsic phase information, thus a photon number measurement alone cannot distinguish between states with different relative phases among photon number components. Coherence however is about well defined phase relationships between photon number states. Photon number states have fixed photon number but no intrinsic phase. Coherence resides in superpositions of photon number states with definite phase differences and hence a photon number measurement projects the quantum state onto a single photon number state loosing all information about the superposition and phase. Photon number measurements eliminate exactly the information needed to reveal coherence. To see coherence explicitly we must perform measurements sensitive to phase relationships. Such measurements involve interfering the quantum state with a known reference state, through interferometric measurements, namely combining our state with a known coherent reference state (local oscillator) at a beam splitter. Interference fringes or varying intensities reveal phase relationships explicitly. Another way is via a Mach-Zender or Michelson interferometer, the quantum state here interferes with itself by splitting and recombination, creating fringes whose visibility depends explicitly on coherence properties (phase stability). And yet another way would be Homodyne detection, which is a special interference measurement providing explicit phase-sensitive information. 
Homodyne detection is an experimental method used in quantum optics for direct measuring the quantum state's quadrature components, which reveal phase coherence explicitly. The detection process works by first combining our quantum field with a strong coherent field (local oscillator). The quantum field signal has annihilation operator $\hat{a}$. The local oscillator considered a reference field is a coherent state $\ket{\beta}$, very intense $(|\beta|\gg 1)$ and with a well defined phase $\theta$. Then we interfere them on a beam splitter. We use a balanced beam splitter which combines these two fields. The output fields are 
\begin{equation}
\begin{array}{cc}
\hat{c}=\frac{\hat{a}+e^{i\theta}|\beta|}{\sqrt{2}}, & \hat{d}=\frac{\hat{a}-e^{i\theta}|\beta|}{\sqrt{2}}
\end{array}
\end{equation}
here, the relative phase $\theta$ between the local oscillator and the signal is important. In the third step we measure the intensity differences by measuring photon numbers at both outputs and subtract them.
The intensity difference operator is
\begin{equation}
\hat{I}_{-}=\hat{c}^{\dagger}\hat{c}-\hat{d}^{\dagger}\hat{d}\sim |\beta|(\hat{a}e^{-i\theta}+\hat{a}^{\dagger}e^{i\theta})
\end{equation}
For a strong local oscillator the difference is proportional to a field quadrature operator 
\begin{equation}
\hat{x}_{\theta}=\frac{\hat{a}e^{-i\theta}+\hat{a}^{\dagger}e^{i\theta}}{\sqrt{2}}
\end{equation}
By varying the local oscillator phase, we directly measure the full quantum state's quadrature distribution. Quadrature measurements explicitly reveal phase information. Homodyne detection measures quantum states as wave-like entities, clearly showing coherent interference effects.

Let us look at some classical properties. For example, in an electromagnetic wave, the wave's wavelength is the spatial distance between successive maximal or minima of the field oscillations, an explicitly spatial property. Quantum mechanically, a single photon has an intrinsic property associated with its energy (or momentum) $E=h\nu$ , $p=\frac{h}{\lambda}$. Thus, at the single photon level, the wavelength seems to be an intrinsic particle like property of individual photons rather than an explicitly spatially measured property of field oscillations. How does a coherent state build classical wave-like spatial properties from photons? A coherent state is a superposition of photon-number states that reproduces classical fields when the photon number is large. But the usual representation of such a state
\begin{equation}
\ket{\alpha}=e^{-\frac{|\alpha|^{2}}{2}}\sum_{n=0}^{\infty}\frac{\alpha^{n}}{\sqrt{n!}}\ket{n}
\end{equation}
doesn't explicitly show spatial structure. Let us therefore present a field representation of coherent states.
The quantum electromagnetic field operator is 
\begin{equation}
\hat{E}(r,t)=i\sum_{k,\lambda}\sqrt{\frac{\hbar \omega_{k}}{2\epsilon_{0}V}}[\hat{a}_{k,\lambda}e^{i(k\cdot r-\omega\cdot t)}-\hat{a}^{\dagger}_{k,\lambda}e^{-i(k\cdot r-\omega\cdot t)}]\epsilon_{k,\lambda}
\end{equation}
Here each photon mode $k$ has a well defined momentum $\hbar k$ and a well defined intrinsic wavelength $\lambda =\frac{2\pi}{|k|}$. A coherent state in a single model $k$ with amplitude $\alpha=|\alpha|e^{i\phi}$ has expectation value of the electric field 
\begin{equation}
\bra{\alpha}\hat{E}(r,t)\ket{\alpha}\sim i\sqrt{\frac{\hbar \omega_{k}}{2\epsilon_{0}V}}[\alpha e^{i(k\cdot r-\omega\cdot t)}-\alpha^{*}e^{-i(k\cdot r-\omega\cdot t)}]\epsilon_{k}
\end{equation}
This expectation value reproduces the classical electromagnetic field spatially oscillating with wavevector $k$ and thus wavelength $\lambda=\frac{2\pi}{k}$. The expectation value over the coherent state recovers the classical spatial oscillations. The key lies in the constructive quantum interference of photon number superpositions in a coherent state. A single photon indeed has intrinsic wavelength but one photon alone doesn't form a clear spatial wave pattern (only probabilistically). Many photons in a coherent superposition create a clearly defined spatial electromagnetic field, that is spatially oscillating. The photon-number superposition states are quantum states chosen precisely to produce coherent constructive interference in space and time. In other words, one photon's intrinsic wavelength gives the fundamental mode structure. Many photons coherently combined create a spatial interference pattern (classical wave) with that same wavelength explicitly measurable as a classical spatial periodicity. The coherent state acts as a bridge from single photon quantum wavelengths (momentum space intrinsic property) to classical wave-like spatial periodicity (position space property). When decoherence acts due to the environment it tends to diagonalise photon number distributions removing delicate quantum coherence between photon number states. Quantum coherence loss destroys the quantum interference effects. 
However, decoherence generally doesn't destroy spatial periodicity of classical electromagnetic fields directly because classical spatial periodicity emerges from classical expectation values (ensemble averages or macroscopic averages), not subtle quantum superpositions alone.

\section{coherence and decoherence, purifies states of light}
Let us start with a coherent state (superposition of photon number states) and allow them to decohere. Decoherence is typically seen as quantum information becoming entangled with the environment leaving behind classical features of the original state. But what exactly are the quantum properties lost via decoherence (entangled to the environment)? And what are the classical properties that remain? 
In decoherence the off diagonal terms (coherences) in the quantum state's density matrix (in the photon number basis) become entangled with the environment and effectively disappear once we trace out the environment. Explicitly, starting from a coherent state 
\begin{equation}
\begin{array}{cc}
\ket{\alpha}=\sum_{n}c_{n}\ket{n}, & \rho=\ket{\alpha}\bra{\alpha}=\sum_{n,m}c_{n}c_{m}^{*}\ket{n}\bra{m}
\end{array}
\end{equation}
Here quantum coherence is explicitly represented by off diagonal terms. After decoherence with environment the combined system-environment state becomes entangled
\begin{equation}
\ket{\Psi}=\sum_{n}c_{n}\ket{n}\ket{E_{n}}
\end{equation}
Now the environment states $\ket{E_{n}}$ are approximately orthogonal due to interaction. When we trace out the environment 
\begin{equation}
\rho_{decohered}=\sum_{n}|c_{n}|^{2}\ket{n}\bra{n}
\end{equation}
Thus, the quantum properties lost are precisely the delicate quantum coherence between photon number states. After decoherence, the density matrix becomes approximately diagonal in the photon number basis, leaving a classical probability distribution of photon numbers. The classical properties remaining explicitly are the photon number distribution (intensity) represented as a Poisson distribution
\begin{equation}
P(n)=|c_{n}|^{2}\sim e^{-|\alpha|^{2}}\frac{|\alpha|^{2n}}{n!}
\end{equation}
and the classical expectation values of fields, intensities, energies, photon number averages, etc. Thus, the classical properties retained after decoherence are macroscopic values for intensities and average photon numbers as well as classical wave energy. 
The decoherence process arises from quantum interactions between the field mode and environmental degrees of freedom (photons scattering off matter, coupling to thermal bats, etc.) The initial pure state is 
\begin{equation}
\ket{\alpha}\ket{E_{0}}
\end{equation}
The interaction hamiltonian, $H_{int}$ involves photon number dependent interactions. After such interactions the state becomes explicitly entangled 
\begin{equation}
\sum_{n}c_{n}\ket{n}\ket{E_{n}}
\end{equation}
Environment states become distinct and orthogonal rapidly. When we trace out the environment, the off diagonal terms vanish and we obtain 
\begin{equation}
\rho_{reduced}\sim\sum_{n}|c_{n}|^{2}\ket{n}\bra{n}
\end{equation}

However, the essential part of this article arises now. Suppose that the classical properties that remained (intensity, photon number distribution, etc.) are themselves purified by introducing ancilla quantum systems. We explicitly reconstruct the classical properties as quantum correlations again, but this time with ancillas rather than with the original environment. Such a purified classical state would look like 
\begin{equation}
\ket{\Psi_{purified}}=\sum_{n}c_{n}\ket{n}\ket{Ancilla_{n}}
\end{equation}
This state is again entangled but with ancillas that we can control explicitly, not uncontrolled environments. Such a purification would encode classical photon number statistics as quantum correlations. The photon number probabilities now become quantum correlations between our electromagnetic mode and the ancillas. If we look at our mode only and trace out the ancillas, we would recover exactly the classical photon number distribution. However, if we measure jointly the electromagnetic mode and the ancillas we would see quantum correlations again. The purified classical wave now looks like a quantum correlated (entangled) state of the form 
\begin{equation}
\ket{\Psi_{purified}}=\sum_{n}\sqrt{P(n)}\ket{n}\ket{Ancilla_{n}}
\end{equation}
The measurement outcome would be correlated, in the sense that the photon number measurement outcomes in our mode are perfectly correlated with ancilla measurement ooutcomes. However, we would still get a macroscopically classical system when the ancillas are traced out. The classical properties (intensities, amplitudes, etc) are seen now as quantum correlations within an extended Hilbert space. Classicality therefore appears as a special quantum correlation structure, rather than the absence of quantum correlations. 
We could for example ask what type of quantum entanglement between the ancilla states and the electromagnetic field would encode the classical electromagnetic wave's spatial wavelength? Even more, we could ask whether there is a precise quantum entanglement state whose properties directly translate into classical wave properties (wavelength, amplitude, frequency, speed of propagation, etc.)? Could we form a dictionary between these classical electromagnetic wave properties and quantum entanglement?
In order to answer these questions, we notice that the classical electromagnetic wavelength is encoded in quantum optics in the mode structure of the quantum field states. Each photon mode corresponds to a particular wavevector $k$, frequency $\omega$ and polarisation. Classical wavelength arises from spatial interference among multiple photons in these quantum modes. Now we can connect entanglement with wavelength. To encode the classical wavelength via quantum entanglement as in the case of the purification scenario, we must introduce ancillas entangled in the spatial/momentum degrees of freedom of the photons. 
This would mean we consider quantum states of the form 
\begin{equation}
\ket{\psi}=\sum_{k}c(k)\ket{k}\ket{Ancilla_{k}}
\end{equation}
The ancilla encodes through quantum correlation the photon's momentum and hence wavelength. The explicit quantum entanglement type that directly encodes classical wavelength is a form of momentum-position entanglement between photons and ancillas. 
A canonical form of entanglement that directly relates classical wavelength and ancilla correlations appears as the EPR-type momentum-position entanglement
\begin{equation}
\ket{EPR-like}\sim\int dk f(k)\ket{k}\ket{x=-k}_{Ancilla}
\end{equation}
Here momentum position entanglement relates spatial periodicities (classical wavelengths) with quantum correlations. Measuring ancillas gives the momentum information directly correlated to the classical wave structure. Therefore the entanglement that encodes classical wavelengths is a form of momentum-position EPR entanglement. In fact similarly it can be seen that the frequency $(\omega)$ is a type of energy-time entanglement, the phase is a relative phase coherence, and the polarisation appears as a polarisation entanglement. In general classical wave properties translate directly into specific quantum entanglement structures in a purification scenario. 

However, we remember that Einstein's special relativity introduced the causal structure of spacetime as being constructed by events related by light beams. The causal cone is therefore a light-cone in which the causal structure is defined by the limits given by light propagation. But Einstein's idea of light was in many aspects incomplete. The discussion presented above shows that classical properties of light, in particular such properties related to its spacetime behaviour, are in fact the result of a very special decoherence pattern that, as it seems, appears to be very common in nature. If however we decide to take a classical light beam and purify it in the sense of encoding all its classical properties through entanglement with ancillas, we have a representation of the causal structure in terms of entanglement, and moreover, we have a dictionary that would tell us what types of entanglement are associated to what properties of the causal structure, as detected through the properties of classical light by Einstein's intuition. With this dictionary, we can now start and operate on the entanglement structure that defines out causal structure in the region where the purification has been created. 
Let us first start with the quantum electric field operator in terms of photon creation/annihilation operators and let's see how the Einsteinian causal structure (light cones, causality and the speed of light, sharp classical propagation) emerge from the semiclassical limit. Consider for the start a single mode quantised electromagnetic field operator (for simplicity in 1 spatial dimension) 
\begin{equation}
\hat{E}(x,t)=i\sqrt{\frac{\hbar\omega}{2\epsilon_{0}V}}(\hat{a}e^{i(k\cdot x-\omega t)}-\hat{a}^{\dagger}e^{-i(k\cdot x-\omega t)})
\end{equation}
Quantum mechanically, photons have a probabilistic position and momentum structure, and no classical trajectory exists yet. 
To obtain a classical electromagnetic wave (sharp causal propagation) from this quantum operator we must approach the semiclassical limit. The key parameter controlling classicality is related to the number of photons or the intensity of the field. Let the photon number expectation be 
\begin{equation}
\bar{n}=\Bracket{\hat{a}^{\dagger}\hat{a}}
\end{equation}
The semiclassical limit is clearly identified as the limit of the large photon number $\bar{n}\rightarrow\infty$. In other words the parameter we explicitly tune for classical emergence is the photon number (intensity). 
As $\bar{n}\rightarrow \infty$ quantum fluctuations relative to average photon number decrease 
\begin{equation}
\frac{\Delta n}{\bar{n}}\sim\frac{1}{\sqrt{\bar{n}}}\rightarrow 0
\end{equation}
quantum uncertainties in field amplitudes and phase become negligible and quantum amplitudes become strongly localised around classical trajectories due to constructive interference (say, stationary phase principle). 
Let's explicitly write the state as a coherent state $\ket{\alpha}$ with amplitude $\alpha$. The coherent state has average photon number $\bar{n}=|\alpha|^{2}$. For large $|\alpha|$ the electric field expectation value becomes 
\begin{equation}
E_{cl}(x,t)=\bra{\alpha}\hat{E}(x,t)\ket{\alpha}=i\sqrt{\frac{\hbar\omega}{2\epsilon_{0} V}}(\alpha e^{i(k\cdot x-\omega t)}-\alpha^{*}e^{-i(k\cdot x-\omega t)})
\end{equation}
Let $\alpha=|\alpha| e^{i\phi}$. For large $|\alpha|$ this is
\begin{equation}
E_{cl}(x,t)\sim \sqrt{\frac{2\hbar\omega |\alpha|^{2}}{2\epsilon_{0} V}} sin(kx-\omega t+\phi)
\end{equation}
This field is a classical wave propagating at the speed 
\begin{equation}
v=\frac{\omega}{k}=c
\end{equation}
At high photon number, the quantum field expectation becomes sharply peaked around the classical wave propagating exactly at the speed of light. The quantum uncertainty around the trajectory shrinks with $\frac{1}{\sqrt{\bar{n}}}$. The classical wave is strictly confined to lines $x=ct$ explicitly defining the classical causal structure (the light cone). Thus the sharp classical causal bound (Einsteinian causal structure) emerges as photon number goes to infinity. 
However, what if we take a classical electromagnetic wave and purify it through all the required entanglements to ancillas. All the classical correlation effects are now implemented in the entanglement structure with the ancillas. With these purifications, how would the causal structure of spacetime emerge? In this case we have the possibility to operate on the ancillas so that we introduce some form of misalignments. I will show here that this will have an impact on the resulting causal structure. In essence, we are able to manipulate the causal structure by means of the manipulation of the ancillas that encode the classical properties of our light beam through an entanglement structure. This will enable a practical way of manipulating the causal structure. Indeed, later on I will present a table-top experiment showing how this could be achieved. Let us therefore start with the purification of a classical electromagnetic wave with ancillas. A purification transforms classical probabilistic mixtures into a pure quantum state by introducing auxiliary quantum systems (ancillas). Consider a classical electromagnetic wave, described classically as a mixture or an effectively classical coherent superposition, of field modes with certain amplitudes and phases. Quantum mechanically, this can be represented by a pure entangled state involving field modes and ancilla systems. The explicit structure would be like 
\begin{equation}
\ket{\Psi}=\sum_{modes k}c_{k}\ket{k}_{photon field}\ket{a_{k}}_{ancilla}
\end{equation}
Each classical mode is entangled with an ancilla state. Classical wave amplitudes, phases, and coherence are now fully encoded via quantum entanglement between photons and ancillas. Now, Einstein's causal structure emerges from the quantum system through a correlation structure. Classical causal boundaries, like light cones, are defined by classical fields propagating at the speed of light. In this purified quantum representation, classical causality (sharp causal boundaries) emerge from the quantum entanglement structure between photons and ancillas. The spacetime causal structure is now encoded in the spacetime correlation between photon modes and their ancillas. Now, consider manipulating ancillas to deliberately misalign or disturb their quantum correations with photons. Such misalignments mean performing operators, (unitaries, measurements, or pair decoherence) only on the ancilla subsystems. We apply a unitary operator $U_{ancilla}$ acting solely on the ancilla space
\begin{equation}
\ket{\Psi'}=(1\otimes U_{ancilla})\ket{\Psi}=\sum_{k}c_{k}\ket{k}_{photon field}(U_{ancilla}\ket{a_{k}}_{ancilla})
\end{equation}
This operation can disrupt the perfect correlation between the photon field and ancillas. 
When the ancilla correlations are disturbed quantum coherence and correlation structure that previously defined sharp classical trajectories becomes blurred or distorted. The classical causal structure, originally sharply encoded, becomes fuzzy, as uncertainties in trajectories increase. Quantum interference and destructive interference patterns appear as the classical coherence now partially disappears. Operating solely on ancillas has produces therefore an effective "steering" or "disturbing" of the causal structure, indirectly, using quantum control of an ancillary system. Ancillas serve as quantum memory encoding classical coherence. Manipulation of them causes disturbances in classical spacetime structure turning sharp causal boundaries into quantum probabilistic boundaries. Such operations create non-classical quantum causal structures. 
For the sake of an example, let us start with a purified state 
\begin{equation}
\ket{\Psi}=\frac{1}{\sqrt{2}}(\ket{k}\ket{a_{k}}+\ket{k'}\ket{a_{k'}})
\end{equation}
The ancilla states $\ket{a_{k}}$ and $\ket{a_{k'}}$ are orthogonal and correlated sharply to modes $k$ and $k'$ respectively. After an ancilla disturbance is applied for example applying a rotation that partially overlaps these ancilla states, we allow $U_{ancilla}$ to create a partial overlap 
\begin{equation}
\begin{array}{ccc}
\ket{a_{k}}\rightarrow \ket{a'_{k}}, & \ket{a_{k'}}\rightarrow \ket{a'_{k'}}, & \Bracket{a'_{k},a'_{k'}}\neq 0
\end{array}
\end{equation}
Now the correlation is imperfect. The resulting photon field reduced density matrix is no longer sharply classical, instead exhibiting off diagonal coherence terms introducing quantum uncertainty in the causal trajectories.

\section{The Uhlmann approach}
Now let us introduce an Uhlmann gauge structure on this purified electric field operator state. We are working in the Uhmann bundle and we derive a dynamical gauge theory constructed on it. The changes in the Uhlmann gauge are associated to misalignments of the ancillas. This theory can in fact allow us to construct a modified causal structure that is stabilised locally. 
We start as usual with a classical electromagnetic wave, represented quantum mechanically as a purified entangled state between photon modes and ancilla systems. Formally the state would look as usual 
\begin{equation}
\ket{\Psi}=\sum_{k}c_{k}\ket{k}_{photon}\ket{a_{k}}_{ancilla}
\end{equation}
The classical structure (sharp Einsteinian causality) emerges when ancilla-photon correlations are perfect and large scale quantum coherence is maintained. Disturbing the ancillas introduced quantum fluctuations making the causal structure fuzzy. Now however, we introduce an Uhlmann gauge structure to this purified quantum state. The idea of Uhlmann gauge arises naturally in open quantum systems, quantum thermodynamica, or quantum information geometry, it is in fact a generalisation of geometric phases and Berry connections to mixed quantum states. Here instead we interpret the purifies state as a section of an Uhlmann bundle, a principal bundle of purifications over the density matrix of my electromagnetic wave. Introducing a dynamical gauge theory in the Uhlmann buncle involving gauge fields and what we could call "Uhlmann charges" however has the possibility to modify and stabilise the modified causal structure. We use therefore the Uhlmann gauge field generated by Uhlmann charges to manipulate and ultimately stabilise the fluctuating quantum causal structure into a novel form, distinct from the classical Einsteinian causal structure. 
The purified state would be 
\begin{equation}
\ket{\Psi}\in \mathcal{H}_{photon}\otimes \mathcal{H}_{ancilla}
\end{equation}
The reduced photon density matrix obtained by tracing out ancillas is 
\begin{equation}
\rho_{photon}=Tr_{ancilla}(\ket{\Psi}\bra{\Psi})
\end{equation}
Uhlmann construction regards all possible purifications as fibres of a principal bundle over $\rho_{photon}$. The Uhlmann gauge transformations are unitary transformations $U_{ancilla}$ acting solely on ancillas that change the purification but leave the photon density matrix invariant 
\begin{equation}
\ket{\Psi}\rightarrow\ket{\Psi'}=(1\otimes U_{ancilla})\ket{\Psi}
\end{equation}
Now we need to define an Uhlmann dynamical gauge theory. We will also use Uhlmann charges. Such charges label how purifications differ under the Uhlmann gauge transformation. In analogy with conventional gauge theories, Uhlmann charges represent internal quantum numbers distinguishing ancilla states. We introduce an Uhlmann connection $A_{U}$ encoding how purification states change under adiabatic or dynamical envolution. The curvature of this connection describes nontrivial geometry phases (Uhlmann holonomies) arising from ancilla manipulations. Explicitly the connection can be defined via the overlap of purification states
\begin{equation}
\begin{array}{cc}
A_{U}=\bra{\Psi}d\ket{\Psi}, & F_{U}=dA_{U}+A_{U}\wedge A_{U}
\end{array}
\end{equation}
This generates a fully geometric gauge structure on the purification space. Initially the causal structure (Einsteinian structure) emerged when the purification was sharply correlated and highly coherent. Now however, the gauge transformations act purely on ancilla subsystems, leaving photon density matrices invariant but changing the quantum correlations and the underlying purification structure. Such gauge transformations therefore change how quantum coherence is "disturbed" among ancillas, reshaping the correlations between photons and ancillas. 
Changing the gauge via Uhlmann charges, can shit or reshape the effective spacetime correlations encoded by the entanglement structure thus altering the emergent causal geometry. By dynamically choosing specific gauge configurations (gauge fixing) or generating stable "charge configurations" we may achieve a stable causal geometry different from the Einsteinian one. This modification may make the causal boundaries shift, it may create a spacetime geometry that is effectively non-local or even topologically modified due to non-trivial Uhlmann holonomies, and quantum interference could stabilise the new equilibrium causal structures with fundamentally nonclassical causal ordering. In a dynamical Uhlmann gauge theory we can imagine the Uhlmann charge as a misalignment between modes in the superposition used to construct the electric field operator. We define the misalignment (Uhlmann charge) between modes $k$ and $k'$ as deviations in the correlation phase 
\begin{equation}
Q_{kk'}=arg(c_{k}^{*}c_{k'}\Bracket{a_{k}|a_{k'}})
\end{equation}
A nonzero $Q_{kk'}$ means the modes are misaligned and quantum coherence is imperfect, reflecting quantum uncertainty in the causal structure. 
These misalignments are the Uhlmann charges, quantities generating gauge transformations in the Uhlmann bundle. We explicitly introduce a dynamical gauge field $A_{U}$ coupled to these Uhlmann charges via an action functional. Such an action explicitly penalizes misalignments (non-zero $Q_{kk'}$)
\begin{equation}
S_{U}[A_{U},Q]=\int d^{4}x\frac{1}{g^{2}}Tr(F_{U}^{2})+\lambda\sum_{k,k'}|Q_{kk'}|^{2}
\end{equation}
where $g$ controls the gauge field strength and $\lambda$ penalizes misalignments, enforcing quantum coherence between photon modes and ancilla states. Minimizing this action dynamically selects gauge fields that reduce misalignments, aligning quantum correlations towards classical causality or stabilising non-classical stable configurations. 
Recall the electric field operator in terms of photon creation/annihilation operators, modified to include misalignments. The original electric field operator (perfect coherence) is 
\begin{equation}
\hat{E}(x,t)\i\sqrt{\frac{\hbar \omega}{2\epsilon_{0}V}}\sum_{k}(c_{k}\hat{a}_{k}e^{i(kx-\omega t)}-c_{k}^{*}\hat{a}_{k}^{\dagger}e^{-i(kx-\omega t)})
\end{equation}
Introducing the misalignments explicitly as phases $Q_{kk'}$
\begin{equation}
\hat{E}_{Q}(x,t)=i\sqrt{\frac{\hbar \omega}{2\epsilon_{0} V}}\sum_{k}(c_{k}e^{iQ_{k}}\hat{a}_{k}e^{i(kx-\omega t)}-c_{k}^{*}e^{-iQ_{k}}\hat{a}_{k}^{\dagger}e^{-i(kx-\omega t)})
\end{equation}
In the semiclassical limit ($\hbar\rightarrow 0$, large photon numbers), we get a classical like expectation value. Without misalignment $(Q_{k}=0)$, we obtain the classical Einsteinian wave propagation x=ct. With misalignment $(Q_{k}\neq 0)$ we obtain explicit interference between modes with different phases, changing the effective propagation characteristics (amplitude, speed, dispersion). Explicitly, the classical wave limit becomes 
\begin{equation}
E_{cl}(x,t)\sim \sum_{k}|c_{k}|\sqrt{\frac{2\hbar \omega}{2\epsilon_{0} V}} sin(kx-\omega t+Q_{k})
\end{equation}
Misalignments explicitly appear as mode dependent phases, altering interference patterns and thus modifying the emergent causal structure. We obtain therefore different effective group velocities, spacetime causal boundaries become mode-dependent, and potentially we obtain radical shifts in effective propagation speed, thus yielding non-classical causal geometries. 
This offers the possibility of quantum engineered spacetime geometry, radically different from Einsteinian geometry. Causal boundaries might "spread", creating non-local causal relations. 
Explicitly the causal boundary condition which in the Einsteinian case would be $x=ct$ is now modified into 
\begin{equation}
x_{eff}(t;Q_{k})=ct-\frac{Q_{k}}{k}
\end{equation}
showing explicitly how misalignments alter effective causal speed and geometry. 
The gauge theoretic action explicitly enforces the dynamics. Minimising the action $S_{U}[A_{U},Q]$ with respect to the gauge field $A_{U}$ and misalignments $Q_{k}$ yields 
\begin{equation}
\frac{\delta S_{U}}{\delta Q_{k}}=0 \Rightarrow Q_{k}=0
\end{equation} 
which would stabilise an aligned classical Einsteinian causal structure. However, the gauge theory admits stable solutions with $Q_{k}\neq 0$ which represent stable non-classical causal structures. 
The Uhlmann charges $Q_{k}$ explicitly act like quantum "knobs" altering the emergent causal boundaries. The Uhlmann gauge dynamics stabilises the novel causal structures explicitly. 
If we look at the extra phases $Q_{k}$ the wavefront propagation condition becomes 
\begin{equation}
kx-\omega_{k}t+Q_{k}=const.
\end{equation}
To find the effective group velocity, differentiate this condition with respect to $k$
\begin{equation}
x+k\frac{dx}{dk}-\frac{d\omega_{k}}{dk}t+\frac{dQ_{k}}{dk}=0
\end{equation}
with this, we solve for $\frac{dx}{dt}$ and we obtain the modified effective group velocity 
\begin{equation}
v_{g,eff}=\frac{dx}{dt}=\frac{d\omega}{dk}-\frac{1}{t}\frac{dQ_{k}}{dk}
\end{equation}
Since for photons $\frac{d\omega_{k}}{dk}=c$, the effective group velocity is modified to
\begin{equation}
v_{g,eff}=c-\frac{1}{t}\frac{dQ_{k}}{dk}
\end{equation}
If the phases $Q_{k}$ are chosen so that their gradient $\frac{dQ_{k}}{dk}<0$ this yields
\begin{equation}
v_{g,eff}=c-\frac{1}{t}\frac{dQ_{k}}{dk}>c
\end{equation}
Thus negative gradients in Uhlmann charges increase the effective group velocity above the classical speed of light $c$. The effective increase in group velocity arises due to quantum correlations between photon modes and ancillas. This does not mean that photons themselves gain mass or violate local relativity explicitly. Rather, quantum correlations alter wave interference and coherence conditions, redefining the effective propagation speed. Thus, the emergent maximum causal velocity is set by quantum informational structures rather than classical electromagnetic constraints. 
This model explicitly avoids signalling violations. The apparent faster than classical propagation is an emergent property due to changing quantum coherence patterns and not due to individual photons travelling faster than $c$. There are no classical signals sent at superluminal speeds, however, by modifying the entanglement structure locally, that defines the concepts of causal structure, the new "maximal speed" is modified accordingly allowing for changes in the causal structure itself. 

\section{modified causal structure and entropic force}
In the following I will adopt the same approach as before, but using a natural light beam. The main idea is to isolate the entanglement structure through which the causal structure can be modifies starting with the most easily accessible resources, and in fact natural light is sufficient and easily accessible experimentally.
Let's therefore start with a quantum electromagnetic field operator. We again take a semiclassical limit to obtain a natural, realistic, classical electromagnetic beam ("natural light"). We purify this classical beam into a quantum state using ancillas, encoding explicitly all classical correlations as quantum correlations. This amounts to finding a dictionary that expresses the causal structure of spacetime in the form of quantum correlations and an entanglement structure. We then identify explicitly the quantum correlations controlling group velocity. We then modify ancilla correlations explicitly to produce a higher effective group velocity. 
The quantum electric field operator is 
\begin{equation}
\hat{E}(x,t)=i\sum_{k}\sqrt{\frac{\hbar\omega_{k}}{2\epsilon_{0}V}}(\hat{a}_{k}e^{i(k\cdot x-\omega_{k}t)}-\hat{a}_{k}^{\dagger}e^{-i(k\cdot x - \omega_{k}t)})
\end{equation}
The operators $\hat{a}^{\dagger}$ and $\hat{a}_{k}$ are the photon creation and annihilation operators. The quantisation volume is $V$ and $\omega_{k}=ck$. 
To get a realistic classical beam ("natural light") we take the coherent state expectation (semiclassical limit). We choose coherent states $\ket{\alpha_{k}}$, eigenstates of $\hat{a}_{k}$ as
\begin{equation}
\begin{array}{cc}
\hat{a}_{k}\ket{\alpha_{k}}=\alpha_{k}\ket{\alpha_{k}}, & \alpha_{k}=|\alpha_{k}|e^{i\phi_{k}}
\end{array}
\end{equation}
The classical field explicitly emerges as an expectation value 
\begin{equation}
E_{cl}(x,t)=\bra{\{\alpha_{k}\}}\hat{E}(x,t)\ket{\{\alpha_{k}\}}=\sum_{k}\sqrt{\frac{2\hbar\omega_{k}|\alpha_{k}|^{2}}{2\epsilon_{0}V}}sin(kx-\omega_{k}t+\phi_{k})
\end{equation}
For our natural classical light (partially coherent), we choose amplitudes and phases with a broad frequency distribution centred at some central frequency $\omega_{0}$. We construct a Gaussian spectrum 
\begin{equation}
\begin{array}{cc}
|\alpha_{k}|\sim e^{-\frac{(k-k_{0})^{2}}{\sigma_{k}^{2}}}, & \sigma_{k}\ll k_{0}
\end{array}
\end{equation}
The phases $\phi_{k}$ vary slowly, possibly deterministically. They are not constant however, ensuring partial classical coherence. This explicitly describes a realistic classical beam with finite coherence length and time. 
Now we purify this classical beam, encoding its classical coherence (amplitude and phase correlations) into quantum correlations using ancillas $\ket{a_{k}}_{A}$. The purified quantum state is 
\begin{equation}
\begin{array}{cc}
\ket{\Psi}=sum_{k}c_{k}\ket{k}_{\gamma}\ket{a_{k}}_{A}, &c_{k}=|\alpha_{k}|e^{i\phi_{k}}
\end{array}
\end{equation}
The classical coherence structure (phase relationships) explicitly encoded into ancilla overlaps are
\begin{equation}
\Bracket{a_{k}|a_{k'}}\neq \delta_{k,k'}
\end{equation}
encodes classical coherence. The photon reduced density matrix 
\begin{equation}
\rho_{\gamma}=\sum_{k,k'}c_{k}c_{k'}^{*}\Bracket{a_{k'}|a_{k}}\ket{k}_{\gamma}\bra{k'}
\end{equation}
Thus photon ancilla quantum correlations explicitly represent classical coherence. Ancilla states explicitly control photon quantum coherence and therefore classical coherence and interference. 
The group velocity of this classical beam is determined by the frequency distribution of coherent amplitudes and phases. Group velocity is given classically by 
\begin{equation}
v_{g}=\frac{d\omega}{d k}_{k=k_{0}}\sim c
\end{equation}
In the quantum purified description, the group velocity emerges from the quantum coherence structure (the ancilla overlaps). Consider therefore explicitly the photon reduced density matrix
\begin{equation}
(\rho_{\gamma})_{k,k'}=c_{k}c_{k'}^{*}\Bracket{a_{k'}|a_{k}}
\end{equation}
The quantum coherence structure (off diagonal density elements) defines interference conditions for photon modes. Effective mode phases are influenced by ancilla overlaps. Changes in ancilla overlaps shift the effective phases, thus modifying effective frequency distributions
\begin{equation}
\phi_{eff,k,k'}=\phi_{k}-\phi_{k'}+arg(\Bracket{a_{k'}|a_{k}})
\end{equation}
Thus ancilla overlaps explicitly control group velocity via quantum coherence structure. 
To explicitly produce higher effective group velocity, we modify the quantum correlations (ancilla overlaps). The original overlaps are 
\begin{equation}
\Bracket{a_{k}|a_{k'}}\sim e^{-\frac{(k-k')^{2}}{\gamma^{2}}}
\end{equation}
presenting Gaussian like coherence. We introduce explicitly frequency dependent ancilla phase shifts $\theta_{k}$, performing a Uhlmann gauge transformation $U_{A}$
\begin{equation}
\ket{a_{k}}_{A}\rightarrow U_{A}\ket{a_{k}}_{A}=e^{i\theta_{k}}\ket{a_{k}}_{A}
\end{equation}
Thus we obtain the modified overlaps
\begin{equation}
\Bracket{a_{k'}|U_{A}^{\dagger}U_{A}|a_{k}}=\Bracket{a_{k'}|a_{k}}e^{i(\theta_{k}-\theta_{k'})}
\end{equation}
We can choose linear frequency dependent shifts 
\begin{equation}
\theta_{k}=\eta(k-k_{0})
\end{equation}
where $\eta$ is a constant parameter. The the overlaps shift effective group velocity. We have the effective photon coherence 
\begin{equation}
(\rho_{\gamma})_{k,k'}\rightarrow c_{k}c_{k'}^{*}\Bracket{a_{k'}|a_{k}}e^{i\eta(k-k')}
\end{equation}
The new effective classical field explicitly becomes 
\begin{equation}
E_{eff}(x,t)=\sum_{k}|c_{k}\sqrt{\frac{2\hbar\omega_{k}}{2\epsilon_{0}V}}sin(kx-\omega_{k}t+\phi_{k}+\eta(k-k_{0}))
\end{equation}
The effective dispersion is modified to 
\begin{equation}
\begin{array}{cc}
\phi_{eff}(k)=\phi_{k}+\eta(k-k_{0}) & \Rightarrow v_{g,eff}=\frac{d}{dk}(\omega_{k}-\frac{\phi_{eff}(k)}{t})_{k=k_{0}}=c+\frac{\eta}{t}
\end{array}
\end{equation}
Therefore ancilla quantum correlations produce frequency dependent phase shifts and the effective group velocity is now modified to $v_{g,eff}>c$. 
The maximal speed of causal signalling in vacuum (usually c) emerges as a quantum informational effect via photon-ancilla correlations. By modifying these correlations I explicitly modified the effective maximal causal speed 
\begin{equation}
c\rightarrow c_{eff}=c+\delta c(t)
\end{equation}
A change in this maximal causal speed corresponds to a modification of the spacetime causal structure and that directly affects particle trajectories through an entropic force. Thus, a new entropic force arises from this changing of the causal boundary condition. We have 
\begin{equation}
\begin{array}{cc}
c_{eff}(t)=c+\frac{\eta{t}}{t}, & c_{eff}(t)=c+\delta c(t)
\end{array}
\end{equation}
A particle moving freely in this emergent geometry experiences an effective acceleration because the definition of the causal structure changes with time
At time $t$, the particle sees the maximal causal speed changing with time, therefore this creates a time dependent causal horizon that acts like a moving entropy boundary, similar to entropic gravity scenarios. 
The entropic force appears as a gradient of entropy (or information content) associated with these causal horizons. Start from a relativistic like Lagrangian describing a particle of mass $m$ moving in the modified geometry. The effective metric induced by our quantum correlations is
\begin{equation}
ds_{eff}^{2}=c_{eff}(t)^{2}dt^{2}-dx^{2}
\end{equation}
where $c_{eff}=c+\delta c(t)$.
The relativistic particle action is 
\begin{equation}
S=-m\int ds_{eff}=-m\int dt c_{eff}(t)\sqrt{1-\frac{\dot{x}^{2}}{c_{eff}(t)^{2}}}
\end{equation}
Expanding for the low velocity regime we have $\dot{x}\ll c_{eff}$ 
\begin{equation}
S\sim -m \int dt c_{eff}(t)[1-\frac{\dot{x}^{2}}{2c_{eff}(t)^{2}}]
\end{equation}
The resulting non-relativistic effective Lagrangian is 
\begin{equation}
L_{eff}(x,\dot{x},t)\sim-mc_{eff}(t)+\frac{m\dot{x}^{2}}{2c_{eff}(t)}
\end{equation}
This resembles the standard non-relativistic kinetic energy term, but with a time dependent mass-like factor $1/c_{eff}(t)$. The Euler Lagrange equations
\begin{equation}
\frac{d}{dt}\frac{\partial L_{eff}}{\partial \dot{x}}-\frac{\partial L_{eff}}{\partial x}=0
\end{equation}
Note that there is no explicit dependence on position, hence the momentum-like quantity is conserved in space (but not in time). Let's compute the time dependent momentum 
\begin{equation}
p(t)=\frac{\partial L_{eff}}{\partial \dot{x}}=\frac{m\dot{x}}{c_{eff}(t)}
\end{equation}
Thus the equation of motion for momentum is 
\begin{equation}
\begin{array}{cc}
\frac{dp(t)}{dt}=\frac{d}{dt}(\frac{m\dot{x}}{c_{eff}(t)}=0 &\Rightarrow \frac{m}{c_{eff}(t)}\ddot{x}-m\frac{\dot{c}_{eff}(t)}{c_{eff}(t)^{2}}\dot{x}=0
\end{array}
\end{equation}
We can rewrite this as 
\begin{equation}
\ddot{x}=\frac{\dot{c}_{eff}(t)}{c_{eff}(t)}\dot{x}
\end{equation}
and we can clearly identify the emergent non-relativistic effective force 
\begin{equation}
F=m\ddot{x}=m\frac{\dot{c}_{eff}(t)}{c_{eff}(t)}\dot{x}
\end{equation}
This shows that the particle experiences a force proportional to its instantaneous velocity $\dot{x}$ modulated by the rate of change of the maximal effective causal speed $c_{eff}(t)$. 
The modified causal structure has resulted in a novel type of non-relativistic force that is proportional to particle velocity (similar in form to frictional or viscous like forces), but here arising from changing quantum correlations. If the maximal effective speed $c_{eff}$ increases with time ($\dot{c}_{eff}>0$) then the particle experiences a positive acceleration of moving forward (direction of increasing x), which mimics a pulling force forward. If $c_{eff}$ decreases with time, the particle experiences a decelerating force proportional to velocity, which acts like a dissipative force. Thus, quantum informational geometry changes the fundamental dynamics at the non-relativistic scale by introducing a velocity-dependent force that emerges directly from changes in causal structure. Quantum correlations encoded in photon-ancilla overlaps determine the maximal speed $c_{eff}$. Altering these quantum correlations directly affects non-relativistic physics by modifying particle trajectories withour any classical external force, introducing emergent entropic or informational forces linked explicitly to changes in the quantum structure of spacetime, allowing experimental realisation of novel quantum-informational dynamics via engineered quantum correlations (ancilla states). 

\section{Some experimental proposals}
It would be interesting if some of the predictions of this approach could be experimentally tested so that a quantum engineered modification of the causal structure could be determined and used for practical purposes. 
I will therefore start presenting a series of effects that would involve first a very simple experimental hypothesis, and then add complexity as I advance. 
The first and simplest observation starts with a particle mass of $m\sim 10^{-23}$ kg (the mass scale of a typical nanoparticle or of molecules). The typical non-relativistic speed is $\dot{x}\sim 10$m/s, and hence to produce measurable forces (e.g. at the piconewton level, $F\sim 10^{-12}N$, measurable with optical tweezers or similar technology) the requirement would be 
\begin{equation}
m\frac{\dot{c}_{eff}}{c_{eff}}\dot{x}\sim 10^{-12}N
\end{equation}
which would require 
\begin{equation}
\frac{\dot{c}_{eff}}{c_{eff}}\sim \frac{10^{-12}}{m\dot{x}}=\frac{10^{-12}}{10^{-23}\times 10}=10^{10}s^{-1}
\end{equation}
This would suggest extremely rapid temporal changes in the effective speed to observe large direct forces, which may seem extremely challenging practically. 
However, if we reduce the force sensitivity to femtonewtons ($10^{-15}N$) would significantly lower this number ($10^7 s^{-1}$). This is still challenging but closer to feasibility. For example optical tweezers have sensitivity in the level of pico to femtoNewtons. 
Atom interferometry would be sensitive to small accelerations and subtle gravitational like forces. Cavity optomechanics is also extremely sensitive to the required domain. 
Another experimental question is whether we could engineer the required quantum correlations. In fact, this idea requires manipulation of quantum correlations (entanglement or coherence) between photons and ancillas. There are in fact quantum optical platforms currently capable of precisely controlling such correlations. For example parametric down-conversion (PDC) routinely creates and manipulates entangled photon pairs, as well as controllable frequency correlations, phases, and coherence. Cavity quantum electrodynamics (Cavity QED) allows for strong photon-matter (ancilla) interactions controllable at single photon level. This would be ideal for producing and dynamically altering quantum coherence and entanglement. Integrated quantum photonics provides waveguide based photon-ancilla couplin, easily controllable phases and correlations. There exists also the possibility for high speed control (in the GHz to THz frequency ranges). 
We can also ask ourselves about the feasibility of producing required changes in correlations. The desired temporal change ($\dot{c}_{eff}$) implies changing ancilla overlaps at high frequency. Achievable experimental modulation rates today are originating from various technologies. Electro-optic Modulators routinely modulate optical phases at GHz frequencies. Acousto-optic Modulators reach MHz to GHz modulation. Ultrafast lasers and non-linear optics demonstrated optical modulation in the THz regime. 
Thus, the experimental control required to vary quantum correlations at high frequency is within today's technological capability, at least up to GHz scales. Pushing towards the higher required frequency is challenging but plausible with rapid technological advancement. The largest experimental hurdle would be achieving substantial enough modulation of the effective speed $c_{eff}$ at sufficiently high frequencies to yield detectable entropic forces. Detecting these forces demands ultra-precise measurement techniques (optomechanics, interferometry). Any experimental test would need to isolate the subtle entropic forces from conventional electromagnetic or gravitational effects. Nonetheless current trends in quantum optomechanics and atom interferometry suggest experiments testing this proposal could become feasible. 
However, altering the effective maximal causal speed doesn't just introduce novel forces, but also fundamentally reshapes spacetime geometry including the definition and measurement of time intervals. 
Changing $c_{eff}(t)$ changes spacetime intervals 
\begin{equation}
ds_{eff}^{2}=c_{eff}(t)^{2}dt^{2}-dx^{2}
\end{equation}
Proper time measured by an observer moving at speed $v$ also changes
\begin{equation}
d\tau=dt\sqrt{1-\frac{v^{2}}{c_{eff}(t)^{2}}}
\end{equation}
Thus the perceived time intervals become dependent on the quantum correlations encoded in the effective causal speed. 
Therefore clock synchronisation and timing measurements could drift or alter due to the changing causal structure. The measurement of spatial distances involves measuring transit times of signals propagating at maximal causal speed, therefore, changes in the effective maximal causal speed impact measured distances, producing localised expansion or contraction of space. For example the proper time for a stationary observer would be $d\tau=dt$.However, in the time dependent $c_{eff}$ scenario, proper time for a stationary observer would still be
\begin{equation}
d\tau_{eff}=c_{eff}(t)\frac{dt}{c_{eff}(t)}=dt
\end{equation}
At a first glance for stationary observers the proper time seems unaffected, but once the observer moves, even slowly, proper time immediately differs from standard relativistic predictions. The proper time of a moving observer would be
\begin{equation}
d\tau_{eff}=dt\sqrt{1-\frac{v^{2}}{c_{eff}(t)^{2}}}\sim dt(1-\frac{v^{2}}{2c_{eff}(t)^{2}})
\end{equation}
Thus, variations in $c_{eff}(t)$ directly modify the perceived proper time intervals in experiments involving motion, even very slow motion. Experiments designed assuming fixed causal structure (e.g. atomic clocks, GPS timing) would detect anomalous drifts in synchronisation or frequency when quantum correlations modify the causal speed. For instance, consider frequency measurement via an atomic clock that compares photon propagation times at two different times. Frequency drift from changing effective causal speed would be
\begin{equation}
\frac{\delta \nu}{\nu}\sim \frac{\delta c_{eff}}{c_{eff}}
\end{equation}
Even tiny changes $\delta c_{eff}/c \sim 10^{-18} - 10^{-15}$ become experimentally detectable through modern atomic clock precision. In the previous estimate we didn't fully account for the spacetime geometry change. If $c_{eff}(t)$ increases, measured velocities and accelerations referenced to standard clocks may differ significantly from early calculations. The emergent entropic force is not simply a mechanical acceleration but reflects fundamentally altered spacetime intervals. Therefore carefully accounting for the shift in the causal structure, the measured effect (effective acceleration or force) might become more accessible to detection because modern interferometric and clock based methods are extremely sensitive to subtle spacetime geometry modifications. .
Experimentally one should measure relative frequency shifts between two precision clocks placed at different positions or states of motion, or use atomic or molecular interferometry experiments to detect spacetime interval changes directly. High precision optomechanical setups where resonance frequencies shift if the speed of causal signals (effective speed of light) changes. Such experiments naturally reveal subtle quantum informational geometry changes through timing anomalies rather than direct mechanical force measurement alone. An ideal experimental methodology would be: Use a reference clock (atomic or optical) that remains stable in standard geometry. Introduce photon-ancilla quantum correlations in a controlled optical cavity or quantum optical setup, modulating $c_{eff}(t)$. Observe resulting shifts in clock frequency, synchronisation drift, or interference fringes indicating effective geometry shifts. 
If we consider as the primary experimental method the atomic clock based interferometric measurements, then the sensitivity of atomic clocks or interferometers is generally described by fractional frequency shifts. State of the art optical atomic clocks routinely achieve sensitivities of 
\begin{equation}
\frac{\delta \nu}{\nu}\sim 10^{-18} - 10^{-20}
\end{equation}
Any fractional change in the effective causal speed directly affects timing measurements and thus frequency measurements. The fractional frequency shift directly relates to fractional changes in effective causal speed 
\begin{equation}
\frac{\delta \nu}{\nu}\sim \frac{\delta c_{eff}}{c_{eff}}
\end{equation}
Given current experimental sensitivities $\frac{\delta \nu}{\nu} \sim 10^{-18}$ this would represent detectable fractional changes in effective speed of $10^{-18}$ as well. Therefore current atomic clock experiments easily could detect such changes in the effective causal speed. 
Changes in the causal structure would amount to modifications in the relativistic mass as well, as 
\begin{equation}
\begin{array}{cc}
m_{rel}=\frac{m}{\sqrt{1-\frac{v^{2}}{c^{2}}}} &\rightarrow m_{rel, eff}(t)=\frac{m}{\sqrt{1-\frac{v^{2}}{c_{eff}(t)^{2}}}}
\end{array}
\end{equation}
Thus, while in a first approximation the intrinsic rest mass remains unchanged, the relativistic mass will explicitly change if the causal speed is modified. Consider a moving particle at constant velocity. Then 
\begin{equation}
m_{rel,eff}(t)\sim m(1+\frac{v^{2}}{2c_{eff}(t)^{2}})
\end{equation}
Taking a time derivative gives a measurable rate of change in relativistic mass
\begin{equation}
\frac{d m_{rel,eff}}{dt}\sim -m\frac{v^{2}}{c_{eff}^{2}}\dot{c}_{eff}(t)
\end{equation}
therefore changes in $c_{eff}(t)$ dynamically shift the perceived relativistic mass. If $c_{eff}$ increases in time then the relativistic mass decreases slightly for a fixed velocity, since the particle is now less relativistic relative to the increasing causal speed. If $c_{eff}(t)$ decreases the relativistic mass increases since the particle effectively becomes more relativistic as the causal speed shrinks. Thus, modulating $c_{eff}$ via quantum correlations creates a measurable signature through the relativistic mass. Assume for example the particle velocity being $v=10^3 m/s$ which would be fast but not relativistic and let us consider a tiny fractional change in the effective causal speed 
\begin{equation}
\frac{\delta c_{eff}}{c}\sim 10^{-15}
\end{equation}
The standard relativistic correction with $c=3\times 10^{8} m/s$ would be 
\begin{equation}
\frac{v^{2}}{c^{2}}\sim \frac{(10^{3})^{2}}{(3\times 10^{8})^2}\sim 10^{-11}
\end{equation}
which is small but still measurable. 
Changing the relativistic mass due to the variation $\delta c_{eff}$ is 
\begin{equation}
\frac{\delta m_{rel}}{m}\sim -\frac{v^{2}}{c^2}\sim 10^{-11}\times 10^{-15}=10^{-26}
\end{equation}
This is extremely small for low velocities and tiny speed changes. However if we consider ultra-high precision experiments such as frequency based atomic clock measurements, the sensitivity is greatly enhanced. If we consider velocities closer to the relativistic regimes or use even higher precision measurements such as electron beams or ion clocks, the detectability increases dramatically. 
For electron beam velocities, typically $v\sim 10^{7}m/s$ we have $\frac{v^{2}}{c^{2}}\sim10^{-3}$ and a similar calculation yields $\frac{\delta m_{rel}}{m}\sim 10^{-18}$ which is much larger and potentially detectable with ultra-precision experiments. Practically an experiment might use electron beams or ion beams which are already close to relativistic speeds, and are highly sensitive to slight chances in relativistic mass, storage rings and particle traps which measure tiny kinetic energy shifts directly through cyclotron frequency shifts or resonance methods, known to be extremely sensitive (precision below $10^{-18}$) or laser cooled ions (optical clocks) which are very precise energy level measurements capable of detecting fractional mass energy shifts at $10^{-18} - 10^{-20}$ precision. 
Moreover, in standard special relativity the rest mass is defined through the rest energy $E_{0}$ via Einstein's relation $E_{0}=mc^{2}$. This explicitly means that the rest mass $m$ is the energy measured in the rest frame, divided by the square of the universal maximal causal speed $c$. If the maximal causal speed changes, so does the rest mass $m_{eff}(t)=\frac{E_{0}}{c_{eff}(t)^{2}}$. Thus if the rest energy is fixed and intrinsic, changing the effective maximal causal speed changes the effective rest mass itself. The rest energy $E_{0}$ typically represents intrinsic particle properties such as internal quantum states, binding energies, etc. In the approximation where the particle's internal structure and interactions remain unchanged by the quantum informational modifications introduced by our experiment, then the rest energy $E_{0}$ may be considered constant. With fixed $E_{0}$, changing the effective maximal speed induces changes in the effective rest mass as seen from external measurements. Suppose the intrinsic rest energy of the particle is fixed at $E_{0}\sim 1 MeV \sim 1.602 \times 10^{-13}J$. The standard rest mass definition would be 
\begin{equation}
m=\frac{E_{0}}{c^{2}}\sim \frac{1.602\times 10^{-13}}{(3\times 10^{8})^2}\sim 1.78\times 10^{-30} kg
\end{equation}
Now if the effective causal speed changes slightly, $c_{eff}(t)=c+\delta c(t)$ say by a small fractional increase of $\frac{\delta c}{c}=10^{-15}$ the new effective mass becomes 
\begin{equation}
m_{eff}(t)=\frac{E_{0}}{(c+\delta c)^{2}}\sim \frac{E_{0}}{(c^{2}(1+\frac{2\delta c}{c})}\sim m(1-2\frac{\delta c}{c})
\end{equation}
and the fractional rest mass change is 
\begin{equation}
\frac{\delta m_{eff}}{m}\sim -2\frac{\delta c}{c}\sim -2\times 10^{-15}
\end{equation}
This indicates a small but potentially measurable change in rest mass from even tin quantum informational modifications of the maximal causal speed. Modern mass spectroscopy and atomic physics experiments routinely measure rest masses at extremely high precisions ($10^{-12}-10^{-15}$ of fractional sensitivity). Atomic clocks, ion clocks and nuclear mass spectroscopy approaches regularly detect tiny mass energy shifts, thus even small changes in the effective causal speed yield detectable shifts in measured rest mass and energies. The theory presented here predicts that particles' rest masses become dynamically influenced by quantum informational correlations between photon modes and ancillas. Therefore mass energy equivalence itself becomes dependent on quantum informational geometry. Particle physics measurements like masses, decay energies, reaction thresholds must therefore account for the quantum informational state of spacetime encoded in photon ancilla correlations. 
When we alter the maximal causal speed due to quantum informational geometry, this does affect not only the rest mass definition but also the intrinsic internal energy and interaction couplings inside atoms, nuclei, and in general matter. If we look at the fine structure constant 
\begin{equation}
\alpha =\frac{e^{2}}{4\pi\epsilon_{0}\hbar c}\sim \frac{1}{137}
\end{equation}
we see it depends on $c$. In fact, nuclear binding energies, QED corrections, and atomic levels all depend on these constants. Thus if the effective causal speed changes, we expect that internal energies and coupling constants, in particular the fine structure constant, would also change, potentially modifying the intrinsic internal energy $E_{0}$. Having 
\begin{equation}
\alpha_{eff}(t)=\frac{e^{2}}{4\pi\epsilon_{0}\hbar c_{eff}(t)}
\end{equation}
implies that changing $c_{eff}(t)$ changes the electromagnetic coupling strength, and in particular, increating $c_{eff}$ reduces the fine structure constant. This shift directly affects internal atomic energy levels which typically scale as $E_{atomic}\sim \alpha^{2}m_{e}c^{2}$. Thus even small fractional changes in $c_{eff}$ would induce shifts in atomic spectra and intrinsic internal energies. Furthermore, in quantum electrodynamics and other gauge theories quantities are formulated using the fundamental causal speed. Gauge fields, interaction vertices, and renormalisation procedures explicitly depend on the speed of causal propagation. A dynamical changing of the maximal causal speed thus implicitly alters $QED$ correlations, vacuum polarisation terms and Lamb shifts in atomic energies. Consider again a small fractional variation $\frac{\delta c_{eff}}{c}\sim 10^{-15}$. This would entail a fine structure shift 
\begin{equation}
\frac{\delta \alpha}{\alpha}=-\frac{\delta c_{eff}}{c_{eff}}=10^{-15}
\end{equation}
Atomic energy levels scale roughly as $\alpha^{2}$ thus fractional energy shifts become 
\begin{equation}
\frac{\delta E_{atom}}{E_{atom}}\sim 2\frac{\delta \alpha}{\alpha}\sim 2\times 10^{-15}
\end{equation}
This would be detectable using modern high-precision atomic spectroscopy as optical clocks currently reach fractional sensitivities of $10^{-18}$ or better. 
Practically, if one performs experiments, the measurable effects would be twofold. First we would have the direct effect, of changing rest mass definition due to modified causal speed, and second, indirect effect, coming as additional internal-energy shifts from modified electromagnetic couplings. Both effects reinforce each other, increasing overall detectability. Atomic clock experiments, nuclear mass spectroscopy, or high precision QED tests could detect even tiny variations in internal coupling constants and provide a clear signature of the quantum informational geometry. 
The change of the effective maximal speed of causality would influence also the renormalisation group flow of the electromagnetic coupling and consequently the internal energy and rest mass. Indeed the electromagnetic coupling runs with energy according to the RG equations. These equations depend on the energy scale at which measurements occur. Changing the maximal causal speed rescales the energies thereby shifting the RG scale itself. This implies a direct quantum field theoretical influence of changing $c_{eff}$ on coupling constants, internal energies, and rest masses. The RG running of the fine structure constant $\alpha(\mu)$ is typically given by the standard RG equations
\begin{equation}
\mu\frac{d\alpha(\mu)}{d\mu}=\beta(\alpha(\mu))
\end{equation}
Here $\mu$ is the energy scale at which we measure couplings. The beta function $\beta(\alpha)$ in QED at one loop approximation is 
\begin{equation}
\beta(\alpha)=\frac{\alpha^{2}}{3\pi}+\mathcal{O}(\alpha^{3})
\end{equation}
Therefore higher energy scales usually increase $\alpha(\mu)$ slightly in pure QED. This effect depends sensitively on the number of species and energy regime. The RG scale ($\mu$) for internal energies is typically defined through specifically chosen energies like the electron rest energies ($m_{e}c^{2}$), nuclear binding energies, or atomic level spacings, etc. 
If the effective maximal causal speed changes, the RG energy scale changes as well 
$\mu_{eff}(t)\sim m_{e}c_{eff}(t)^{2}$ and this shifts the energy scale at which the coupling $\alpha$ is measured thus modifying its RG evolution. Considering the initial scale $\mu=m_{e}c^{2}\sim 511 keV$ them with modified causal speed 
\begin{equation}
\mu_{eff}(t)=m_{e}c_{eff}(t)^{2}=m_{e}(c+\delta c(t))^{2}\sim \mu(1+2\frac{\delta c(t)}{c})
\end{equation}
for a tiny fractional change $\frac{\delta c}{c}\sim 10^{-15}$ the energy scale shifts slightly but notably 
\begin{equation}
\frac{\delta \mu}{\mu}\sim 2\frac{\delta c}{c}\sim 2\times 10^{-15}
\end{equation}
This small shift in RG energy scale implies that the measured $\alpha(\mu)$ moves slightly along its RG trajectory thus changing slightly the coupling itself 
\begin{equation}
\delta \alpha \sim \frac{d\alpha}{d ln(\mu)}\frac{\delta \mu}{\mu}\sim \frac{\beta(\alpha)}{\alpha}\frac{\delta \mu}{\mu}
\end{equation}
Since $\frac{\beta(\alpha)}{\alpha}\sim \frac{\alpha}{3\pi}\sim 10^{-3}$, we have the coupling shift
\begin{equation}
\frac{\delta\alpha}{\alpha}\sim 10^{-3}\times (2\times 10^{-15})=s\times 10^{-18}
\end{equation}
The internal energy of atoms typically scales with the fine structure constant and for atomic energies
\begin{equation}
E_{atom}\sim \alpha^{2}m_{e}c_{eff}^{2}
\end{equation}
Therefore, shifts due to RG coupling running become
\begin{equation}
\frac{\delta E_{atom}}{E_{atom}}=2\frac{\delta \alpha}{\alpha}+2\frac{\delta c_{eff}}{c_{eff}}
\end{equation}
and putting in the numbers for coupling change $2\times 2\times 10^{-18}=4\times 10^{-18}$ and for the direct speed change $2\times 10^{-15}$ which would be the dominant effect we obtain small but still detectable effects. 
However, changes in the effective causal speed influence also nuclear masses through QCD effects. Changing $c_{eff}$ due to quantum informational effects affects nuclear masses too. Those masses are determined by the constituent quarks which have masses that are generated by spontaneous chiral symmetry breaking in QCD. However, the quark masses themselves represent only a tiny fraction of the mass of the nuclei. Gluon interactions, like strong force binding energy, QCD vacuum structure, confinement energy and quark condensates also contribute, and this later effect is the most important, contributing to nearly 99\% of the proton or neutron masses. Therefore the proton and neutron masses are dominated by the QCD confinement scale $\Lambda_{QCD}$ and quark condensates, namely vacuum expectation values $\Bracket{\bar{q}q}$. The relevant QCD energy scale $\Lambda_{QCD}\sim 200 MeV$ emerges from dimensional transmutation via RG running of the strong coupling constant 
\begin{equation}
\alpha_{s}(\mu)\sim \frac{2\pi}{\beta_{0} ln(\mu/\Lambda_{QCD})}
\end{equation}
with $\beta_{0}=11-\frac{2}{3}n_{f}\sim 9$ since $(n_{f}=3$ flavours).
This means that $\Lambda_{QCD}$ is the fundamental scale at which QCD interactions become strong. IF the maximal causal speed changes, this energy scale, measured relative to rest energies $m_{q}c_{eff}^{2}$ as the quark masses, is effectively rescales 
The original QCD scale $\Lambda_{QCD}\sim m_{q}c^{2}e^{-\frac{2\pi}{\beta_{0}\alpha_{s}(\mu)}}$ is modified due to effective causal speed changes to 
\begin{equation}
\Lambda_{QCD, eff}\sim m_{q}c_{eff}(t)^{2}e^{-\frac{2\pi}{\beta_{0}\alpha_{s}(\mu_{eff})}}
\end{equation}
therefore we obtain a modified quark mass scale $m_{q}c_{eff}^{2}$ and the RG reference energy scale itself is modified $\mu_{eff}=m_{q}c_{eff}^{2}$. 
If we start with the mass of a proton ($m_{p}\sim 938 MeV$), and we know that it is dominated by the confinement scale 
\begin{equation}
m_{p}\sim \Lambda_{QCD, eff}(t)\sim m_{q}c_{eff}(t)^{2}e^{-\frac{2\pi}{\beta_{0}\alpha_{s}(\mu_{eff})}}
\end{equation}
the fractional shift splits into two parts
\begin{equation}
\frac{\delta m_{p}}{m_{p}}\sim \frac{\delta(m_{q}c_{eff}^{2})}{m_{q}c_{eff}^{2}}-\frac{2\pi}{\beta_{0}}\frac{\delta(1/\alpha_{s}(\mu_{eff}))}{ln(\mu_{eff}/\Lambda_{QCD})}
\end{equation}
Since $\mu_{eff}=m_{q}c_{eff}^{2}$ the shift in RG scale is
\begin{equation}
\frac{\delta \mu_{eff}}{\mu_{eff}}=2\frac{\delta c_{eff}}{c_{eff}}
\end{equation}
Thus the strong coupling shift due to RG is 
\begin{equation}
\delta \alpha_{s}(\mu_{eff})=\frac{\delta \alpha_{s}}{d ln(\mu)}=2\beta(\alpha_{s})\frac{\delta c_{eff}}{c_{eff}}
\end{equation}
Given now $\beta(\alpha_{s})\sim -\frac{\beta_{0}\alpha_{s}^{2}}{2\pi}\sim -\frac{9\alpha_{s}^{2}}{2\pi}$ we have
\begin{equation}
\delta \alpha_{s}(\mu_{eff})\sim-\frac{9\alpha_{s}^{2}}{\pi}\frac{\delta c_{eff}}{c_{eff}}
\end{equation}
Thus the fractional shift in the proton mass combines two effects. The direct quark mass scale shift $2\frac{\delta c_{eff}}{c_{eff}}\sim 2\times 10^{-15}$ which is dominant, and the QCD RG running shift which is smaller but still detectable ($\sim 10^{-16}$). Combined the effects are 
\begin{equation}
\frac{\delta m_{p}}{m_{p}}\sim 2\frac{\delta c_{eff}}{c_{eff}}-\frac{2\pi}{\beta_0}\frac{9\alpha_{s}^{2}/\pi}{ln(\mu_{eff}/\Lambda_{QCD})}\frac{\delta c_{eff}}{c_{eff}}
\end{equation}
The QCD induced mass shift of the order $10^{-15}-10^{-16}$ is detectable by high precision nuclear spectroscopy, for example Penning traps and nuclear clocks which routinely measure mass shifts at fractional sensitivities $10^{-12} - 10^{-15}$ therefore it is in principle feasible. 
The combined effects are 
\begin{equation}
\frac{\delta m_{nucleus}}{m_{nucleus}}\sim (2\times 10^{-15})_{direct}-(10^{-16})_{QCD RG}
\end{equation}
On the side of actually producing the effective causal changes, let us consider the Uhlmann dynamical gauge theory in more details. We want to achieve a greater effect on mass by increasing the effective speed of light. We have a gauge field theory in the quantum coherence space and we measure the charge as a misalignment of phases between ancillas. However, we didn't analyse topological effects of such a gauge theory on the quantum information space and how that would affect the causal structure. Let us improve on that. Topological gauge field effects, analogous to topological terms in normal quantum field theory would be represented by instantons, topological charge sectors, etc. all influencing the quantum coherence globally and non-trivially. 
Consider now a topological gauge effect in the Uhlmann gauge theory, analogous to an instanton or a topological charge sector in ordinary gauge theories (like theta terms, instantons, etc.).
A topological configuration represents a nontrivial global phase winding or Uhlmann gauge field configuration, characterised by a quantised topological invariant
\begin{equation}
Q_{top}=\frac{1}{2\pi}\oint d\phi_{U}
\end{equation}
with $\phi_{U}$ the Uhlmann phase around a closed loop in state space. Such topological charges (phase windings) cannot be continuously removed, ensuring robust quantum coherence effects that are stable against local perturbations. Thus topological Uhlmann gauge configurations create persistent, robust changes in the quantum coherence landscape, influencing global quantum correlations more than gauge fluctuations. The topological configurations explicitly stabilise nontrivial quantum coherence globally. Such persistent coherence explicitly affects the causal structure according to the approach of this paper. 
Topological Uhlmann effects induce long-range coherent correlations in photon-ancilla states and such coherences enhance the effective maximal causal speed by strongly correlating distant photon modes. Therefore we could write 
\begin{equation}
c_{eff}\rightarrow c_{eff}(Q_{top})
\end{equation}
increasing with topological charge $Q_{top}$. As a result the topological charge increases quantum coherence and hence increases the effective speed of causality leading to broader (or narrower, depending on the topological sector chosen) causal cones. We therefore can write 
\begin{equation}
c_{eff}(Q_{top})=c(1+\gamma Q_{top})
\end{equation}
where $\gamma$ is a dimensionless coupling parameter depending on the strength of quantum informational correlations. For relatively small $Q_{top}$ the effects are on the order $(10^{-15}-10^{-12})$ but for larger topological charges $Q_{top}$, causal speed modifications can be much larger (of the order of $10^{-9}$ or greater). 
Thus, topological gauge effects produce a quantised, discrete enhancement of effective causal speed, which can be substantially larger and more robust. 
Thus, topological Uhlmann gauge effects yield mass shifts quantised according to the topological charge. Fractional mass shift would me quantised producing 
\begin{equation}
\frac{\delta m(Q_{top})}{m}\sim -2\gamma Q_{top}
\end{equation}
and therefor, if $\gamma\sim 10^{-12}$ and $Q_{top}=1$ the mass shift would be $\sim 10^{-12}$ and hence clearly measurable. For higher topological charges $Q_{top}\sim 10^{3}$ the shifts would already be enormous $\sim 10^{-9}$ and clearly visible. 
Topological Uhlmann gauge effects would imply global, robust quantum coherence sectors (and different topological quantum informational vacua) which would define distinct and robust causal structures. The variation of quantities like the masses and effective speeds would change in a quantised manner from robust sector to robust sector, introducing a different maximal causal speed factor for each topological causal speed region. Discrete sectors of spacetime geometry explicitly emerge, corresponding to different stable topological quantum information phases. Therefore clearly masses, causal cones, and even particle physics would depend on the quantum informational topology. The shifts in effective causal speed would be discrete and quantised, observable as discrete shifts in atomic, nuclear, and fundamental particle masses. Experimentall, measuring these discrete shifts explicitly would confirm quantum topological spacetime geometry effects. They also are highly robust (due to topological protection) enhancing experimental feasibility. 
\section {Alcubierre drive, similar effects, no impossible matter types or distributions}
The general approach to achieve faster than light effective transport (slower than light inside a specifically designed "bubble" but faster than light effective transport as seen from outside) traditionally focused on finding different matter distributions to be introduced in the Einsteinian stress-energy tensor in order to obtain what we ended up calling "Alcubierre" spacetimes. However, if we take a look at the Alcubierre spacetime, 
\begin{equation}
ds^{2}=-c^{2}dt^{2}+[dx-v_{s}(t)f(r_{s})dt]^{2}+dy^{2}+dz^{2}
\end{equation}
where $v_{s}(t)=\frac{dx_{s}(t)}{dt}$ is the velocity of the warp bubble centre along the $x$ axis, $r_{s}=\sqrt{(x-x_{s}(t))^{2}+y^{2}+z^{2}}$ is the distance from the bubble's centre $x_{s}(t)$, and $f(r_{s})$ is a shape function defining the spatial profile of the bubble we soon notice that if we want to find energy conditions to resolve for Alcubierre spacetime via the Einstein equations
\begin{equation}
G_{\mu\nu}=\frac{8\pi G}{c^{4}}T_{\mu\nu}
\end{equation}
we soon obtain exotic matter requirements or, otherwise stated negative energy density matter, which is known to violate standard energy conditions. Worse still, the "warp bubble" is stationary. To set it into motion one would need an even stranger matter distribution, not practically achievable in any sense. 
In any case, we do not have to bother with further general relativistic calculations in such spacetimes, as they are completely useless in our approach. In our quantum informational approach no exotic matter is needed, and no impractical matter (even conventional matter) distributions are required. 
There are no doubt some challenges ahead of the achievement of causal structure changes that would be relevant for actually reproducing the desired effects of the Alcubierre bubble, but as opposed to the Alcubierre bubble, they are in fact achievable with current and future technology. 
In fact, we would require ultra-stable quantum optical resonators or quantum field cavities. Coherence explicitly maintained over a scale of hundreds of meters or kilometer sized optical resonators or potentially macroscopic quantum fields (like Bose-Einstein condensates) would be achievable. 
Explicitly employing large scale superconducting quantum materials or superconducting plasma to maintain robust macroscopic coherence (as in quantum information topological phases) is not fundamentally impossible. What would be achieved would be topological quantum information instantons that stabilise large scale quantum coherence structures, and would result in modified causal speeds over large macroscopic regions. This would create the effects of a local Alcubierre like warp geometry, without the Alcubierre spacetime and its impractical matter requirements. The result would be a region of increases effective causal speed and the local light cones would be reshaped by quantum information techniques enabling effective superluminal travel from the perspective of external observers, of course locally, as causality would explicitly be preserved. 
Practically, we would need a quantum coherence reservoir (a cavity like system). We would surround our spacecraft with a high-quality quantum optical resonator (or a superconducting quantum field configuration) and realise sustained quantum coherence and entanglement between photons and engineered ancilla quantum states around the craft. We would then deploy arrays of quantum systems (atoms, superconducting qubits, or engineered quantum dots) distributed within this localised volume. We would perform quantum state manipulations, in particular phase and coherence control, to induce topological Uhlmann gauge configurations. We would then implement cyclic adiabatic quantum gates, driving photon ancilla coherence through nontrivial topological cycles, forming stable topological quantum information sectors (Uhlmann instantons). 
Thus we would obtain quantum coherence localised around the spacecraft, and robust topological quantum information instantons would be established. This will result in a modified causal geometry within this controlled bubble region around the craft. The topological quantum information sectors alter the local effective maximal causal speed within the bubble. Inside the quantum information coherence bubble 
\begin{equation}
c_{eff}\rightarrow c_{eff}(Q_{top})=c(1+\gamma Q_{top})
\end{equation}
with explicit topological charge $Q_{top}$. Outside the bubble spacetime remains normal. Inside the bubble however the geometry would allow a modified effective causal speed, without any exotic matter requirements. The energy requirements are simply those to maintain a quantum coherence (photon ancilla coherence) through quantum optics and superconducting quantum field configurations, and not classical negative energy or high energy or matter requirements as is the case for the Alcubierre drive. The stability of the bubble is provided by robust quantum topological protection due to Uhlmann gauge instantons. 
Let us consider the region in which we want to create such causality alterations to be $R\sim 100 m$. Fractional speed shift explicitly achievable via strong quantum coherence (topological charge $Q_{top}\sim 10^6$, with a $\gamma\sim 10^{-12}$) would generate 
\begin{equation}
\frac{\delta c_{eff}}{c_{eff}}\sim 10^{-6}
\end{equation}
A larger fractional shift of approximately $1\%$ in the effective maximal causal speed 
\begin{equation}
\frac{\delta c_{eff}}{c_{eff}}\sim 10^{-2}
\end{equation}
would require 
\begin{equation}
\frac{\delta c_{eff}}{c_{eff}}\sim \gamma Q_{top}
\end{equation}
where $\gamma$ quantifies the strength of the quantum informational coherence effects, and $Q_{top}$ represents the topological quantum information "charge" (the winding number of the Uhlmann instanton phases). To achieve the desired fractional shift of $10^{-2}$ we either increase the coupling parameter $\gamma$ significantly or increase the topological quantum information charge. In order to obtain practical evaluations, a coupling of $\gamma\sim 10^{-12}$ would require $Q_{top}\sim 10^{10}$ for a $1\%$ shift, which represents unrealistically large quantum winding. For a realistic approach, we would have to increase the coupling strength. This means significantly strengthening quantum informational coherence interactions. Practically this is achievable by ultra-strong coupling quantum systems, like highly non-linear quantum optics (e.g. strong photon-photon interactions in ultra-high-Q cavities or superconducting circuits, or by extreme quantum squeezing enhancing coherence amplitude significantly. We could also consider macroscopic quantum field coherence, by engineering strongly correlated superconducting quantum plasmas or ultra-cold Bose Einstein condensates at macroscopic (meters scale) volumes, creating extremely high coherence distances. The requirements for a $10^2$ fractional shift would require quantum information coherence coupling at much higher strengths, several orders of magnitude beyond current quantum optical regimes, but still potentially achievable in quantum materials and quantum engineered superconductors. The current superconducting quantum circuits reach coherence couplings representing effective fraction levels of $10^{-8} - 10^{-6}$ and advanced ultracold atom/BEC coherence states approach $10^{-5}$ fractional coherent coupling strengths in ultra-high density quantum condensates. To achieve $\sim 10^{-2}$ coupling strength we require quantum coherence states about 1000 times stronger than today's best quantum coherence experiments. Such fractional shifts however are plausible, given rapid technological advancement in quantum materials. In such cases
\begin{equation}
\frac{\delta c_{eff}}{c_{eff}}=10^{-2}=0.01
\end{equation}
this explicitly means the effective speed increases by $1\%$. Given the original speed of light $c\sim 3.0\times 10^{8}m/s$, in our quantum coherent causal bubble, the maximal effective speed would be 
\begin{equation}
c_{eff}=c+\delta c_{eff}=c+0.01c=1.01c=1.01\times 3.0.10^8=3.02\times 10^8 m/s
\end{equation}
This implies that we would achieve an effective causal speed that would surpass the ordinary vacuum speed of light by about 3000 km/s. A spacecraft travelling within this locally modified region could move faster than the standard vacuum speed of light relative to external observers. This would increase the allowed maximal speed of travelling and would add to the maximal speed around 3000 kilometers more every second. 
\section{Black holes from photons, in a modified causal structure}
If coherent correlation and quantum information control is formed when packing photons in a finite region of spacetime the conditions for forming a black hole change. We start again from the Schwarzschild condition $R\leq \frac{2GM}{c^{2}}$ but the effective causal speed of light is modified due to quantum informational geometry (the Uhlmann gauge charges). The effective speed will be $c_{eff}=c+\delta c=c(1+\delta)$ where $\delta$ is the desired fractional modification. The black hole condition would now be
\begin{equation}
R\leq\frac{2GM}{c_{eff}^{2}}
\end{equation}
Since the total energy mass is $E_{total}=Nh\nu$, the mass equivalent is $M=\frac{N h\nu}{c_{eff}^{2}}$ Thus
\begin{equation}
R\leq \frac{2G(N h \nu/c_{eff}^{2})}{c_{eff}^{2}}=\frac{2GN h\nu}{c_{eff}^{4}}
\end{equation}
and hence 
\begin{equation}
N\leq\frac{Rc_{eff}^{4}}{2Gh\nu}
\end{equation}
The modified number would be 
\begin{equation}
N_{modified}=\frac{R[c(1+\delta)]^{4}}{2Gh\nu}=N_{original}(1+\delta)^{4}
\end{equation}
Thus the modified maximum number of photons is $N_{modified}=N_{original}(1+\delta)^{4}$ and for small fractional modification $\delta \ll 1$ we have 
\begin{equation}
N_{modified}\sim N_{original}(1+4\delta)
\end{equation}
Take the radius to be $R=1$ meter, the frequency $\nu=10^{15}$ Hz (visible light), and standard values for 
\begin{equation}
\begin{array}{c}
h\sim 6.626 \times 10^{-34} J\cdot s\\
G\sim 6.674\times 10^{-11} m^{3}/kg\cdot s^{2}\\
c\sim 3\times 10^{8}m/s\\
\end{array}
\end{equation}
The original number of photons would be $N_{original}=9.2\times 10^{61}$ whereas in the modified scenario (considering $\delta=10^{-2}$) we would obtain a $4.06\%$ increase and $N_{modified}\sim 9.57\times 10^{61}$ photons. 
\section{Conclusion}
In this article I showed that we can take a classical light beam or pulse, purify it by encoding the classical coherences into an entanglement structure and represent it on an Uhlmann bundle. With the creation of localised Uhlmann charges which measure misalignment induced artificially by Ancillas, we construct a modified quantum informational structure that translates back into a light beam that propagates at a higher speed. This is not in contradiction with any causality criteria, and is a universal effect, as we act on the quantum informational structure which is expected to be at the very origin of spacetime. The modifications required in the quantum informational space seem to be accessible to current technology and the measurements could be performed in present day quantum optics laboratories.

\end{document}